\documentclass[12pt]{article}

\usepackage{times}

\usepackage{graphicx}% Include figure files
\usepackage{dcolumn}% Align table columns on decimal point
\usepackage{bm}% bold math
\usepackage{bbold}
\usepackage{braket}
\usepackage[ruled,vlined]{algorithm2e} % Algorithms
\usepackage{amsmath}
\usepackage{multibib}
\newcites{Methods}{Methods references}
\usepackage[displaymath]{lineno}
\usepackage[svgnames]{xcolor}
\usepackage{setspace}
\DeclareMathOperator{\tr}{Tr}

\DeclareMathOperator{\real}{Re}

\newcommand{\id}{\mathbb{1}} % unit matrix symbol
\newcommand{\new}[1]{\textcolor{black}{#1}} % To mark changes for the referees

\begin{document}

%\preprint{APS/123-QED}

\title{%Design principle for disordered media with perfect transmission\\or\\Disordered media with perfect transmission for all incoming wave fronts \\ or \\
Customized anti-reflection structure for perfect transmission through complex media}

\author
{Michael Horodynski,$^{1}$ Matthias K\"uhmayer,$^{1}$ Cl{\'e}ment Ferise,$^{2}$ \\ Stefan Rotter,$^{1}$ Matthieu Davy.$^{2}$
\\
\normalsize{$^{1}$Institute for Theoretical Physics, Vienna University of Technology (TU Wien),}\\
\normalsize{A–1040 Vienna, Austria }\\
\normalsize{$^{2}$Univ.\ Rennes, CNRS, IETR (Institut d'{\'E}lectronique et des Technologies du num{\'e}Rique),}\\
\normalsize{UMR–6164, F–35000 Rennes, France}
}

\date{}

\maketitle 

% A brief teaser statement highlighting main result of the paper, understandable by a scientist not in your field, without jargon or abbreviations. This will appear online adjacent to the title and should not repeat phrases already present there. Please keep to under 125 characters and spaces
%\begin{sciabstract} %Currently 125 characters (including spaces)
%    \underline{One-Sentence Summary:} We show that a complex medium can be made fully transmitting to all wave fronts by overlaying it with a complementary medium.
%\end{sciabstract}

%Max 200 words (referenced), currently 197
\noindent \textbf{Getting to grips with the detrimental influence of disordered environments on wave propagation is an interdisciplinary endeavour spanning diverse research areas ranging from telecommunications \cite{basar_wireless_2019} and bio-medical imaging \cite{kubby_wavefront_2019} to seismology \cite{Campillo2003} and material engineering \cite{Chen2010,Molesky2018}. Wavefront shaping techniques are highly promising to overcome the effect of wave scattering as even opaque media feature open channels for which the incident light is fully transmitted \cite{Dorokhov1984,Gerardin2014,Sarma2016,Jeong18}. With this feature being restricted, however, to just a small subset of judiciously engineered states it remains out of reach to render an opaque sample translucent for any incident light field. Here we show that a structureless medium composed of randomly assembled scattering elements can be made fully transmitting to all incoming wavefronts by putting a customized complementary medium in front of it. This special situation is achieved when the reflection matrices of the two media surfaces facing each other satisfy a matrix generalization of the condition for critical coupling. We implement this protocol both numerically and experimentally for the design of electromagnetic waveguides with several dozen scattering elements placed inside of them. The translucent scattering media we introduce here also have the promising property of being able to store incident radiation in their interior for remarkably long times.}

Be it the reduced connectivity to a wireless network, the fading of a radio or the limited line of sight in a foggy environment: in all of these cases the scattering of waves in disordered media leads to interference patterns with a seemingly uncontrollable complexity. To overcome these difficulties, great hopes have been placed on a fundamental and striking result of wave transport theory \cite{Dorokhov1984}, which shows that even opaque media that are highly scattering for incoming waves, feature \textit{open} transmission channels. When suitably excited, light can sneak through these channels even across a complicated maze of disorder with perfect transmission and no back-reflection to the input. While the level of control that is required to engineer such states in practice remains a challenge even for advanced wavefront shaping techniques, remarkable results have been achieved with open transmission channels both in optics and acoustics \cite{Gerardin2014,Sarma2016,Jeong18}. Recently also so-called ``scattering-invariant modes'' have been demonstrated whose transmitted output patterns are the same, irrespective of whether a scattering medium is placed in their way or not \cite{pai_scattering_2021}. %Similar to the open transmission channels, however, the feature of strong transmission remains restricted also here to only a small sub-set of judiciously engineered input light fields.

The fundamental and far-reaching question we address here is, whether an operational procedure can be found such that instead of only a single and specifically designed light field, \new{the energy of} \textit{all} field patterns impinging on a disordered medium gets perfectly transmitted \new{(irrespective of the spatial distribution of the transmitted field)}. Clearly, wavefront shaping at the input of a given medium will not suffice for this challenging goal, since only a fraction of transmission channels are \textit{open}, with all others being \textit{closed}  or having \textit{intermediate} transmission and reflection. %next to the \textit{open} transmission channels there will always also be the \textit{closed} channels and those with \textit{intermediate} transmission and reflection. 
Instead, the only option seems to be the engineering of the disordered medium itself. Indeed, several protocols have recently been put forward for the design of hyper-uniform and non-Hermitian media that feature perfect transmission even for certain disordered internal structures \cite{florescu_designer_2009,leseur_high-density_2016,horsley_spatial_2015,rivet_constant-pressure_2018}. While such strategies may be usable during the fabrication process of materials with desired properties, applying them to already existing disordered media in post-fabrication is extremely challenging, in particular as the internal structure of such media is typically not only unknown but also inaccessible. 

Here, we present a viable strategy for rendering an arbitrary and immutable disordered medium completely transmitting to \textit{all} possible incoming wavefronts by putting a suitably engineered complementary medium right in front of it. % (see illustration in Fig.~\ref{fig:concept}). 
Unlike matching layers that suppress impedance mismatch only in stratified media \cite{Chen2010,Spinelli2012,Im2018}, this prepended ``anti-reflection structure'' is a disordered medium itself, with the special property that it is perfectly matched to the given medium one aims to transmit across. Most importantly, the internal structure of the given medium does not need to be known. Instead, the relevant matching condition solely relies on the given medium's one-sided reflection matrix that is routinely measured experimentally. The remaining task to be solved is to engineer the complementary medium in front such that it has a desired one-sided reflection matrix itself to satisfy the matching condition involving both media.

\begin{figure*}[!b]
\includegraphics[width=0.94\textwidth]{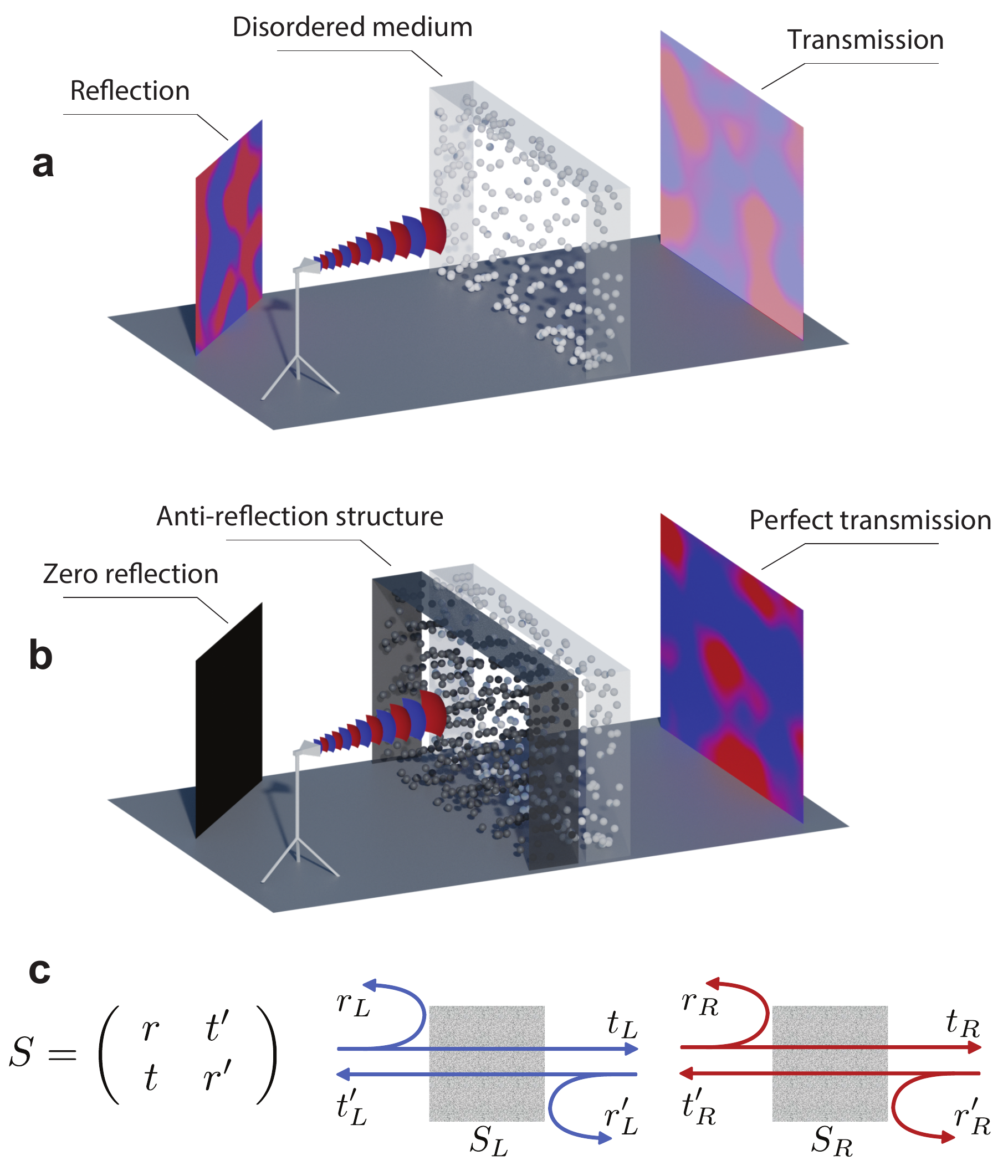}
\caption{\label{fig:concept} \textbf{Illustration of the concept.} \textbf{a}, Transmission of a wave impinging on a disordered medium is typically reduced by multiple scattering. \textbf{b}, By superimposing a customised complementary medium, however, perfect transmission and zero reflection is obtained for any incoming wavefront. \new{\textbf{c}, The $2N\times2N$ scattering matrix $S$ containing $N\times N$ reflection and transmission matrices, can be expressed as a composite structure including the scattering matrices $S_L$ of the optimised anti-reflection structure on the left ($L$) and $S_R$ of the given disordered medium on the right ($R$).}}
\end{figure*}

\section*{Full transmission through disorder}

Our concept is illustrated in Fig.~\ref{fig:concept}: starting point is a given disordered medium of thickness $L$, for which the average transmission of a wavefront impinging from the left typically scales as $\langle T \rangle \sim \ell / L$, where $\ell$ is the mean free path. Correspondingly, when placing a second disordered medium directly to the left of the first one, the transmission through both media would be further reduced as the total scattering region is now increased in length. Here, instead, we aim to achieve the opposite: by designing the medium on the left ($L$) to be complementary to the medium on the right ($R$), the total transmission through the composite structure should reach the maximum possible value as obtained for the case when both media were not even present and all incoming fields pass through the structure without back-reflection. To identify the proper condition for complementarity, consider first the frequency-dependent scattering matrix $S(\nu)$ of the total system. In the absence of absorption, the scattering matrix is unitary as a consequence of flux conservation, $S^\dagger S = \id$. This scattering matrix is composed of two reflection matrices $r$ and $r'$ for incoming flux from the left (unprimed) and right side (primed) of the sample, respectively, and of the corresponding transmission matrices $t$ and $t'=t^T$.  Simple expressions now relate the total scattering matrix $S$ of the composite structure to the scattering matrices $S_L$ and $S_R$ of the left and right medium, respectively \new{(see Fig.~\ref{fig:concept}c for an illustration)}:
\begin{eqnarray}
t = t_R (\id-r_L^\prime r_R)^{-1} t_L \ , \\
r = r_L + t_L^\prime r_R (\id - r_L^\prime r_R)^{-1} t_L \ .
\end{eqnarray}
\noindent Remarkably, using the relations $t_L t_L^\dagger + r_L^\prime r_L^{\prime\ \dagger} = \id$ and $r_L t_L^\dagger + t_L^\prime r_L^{\prime\ \dagger} = 0$ following from the unitarity of $S_L$, we deduce that if the reflection matrix of the complementary medium $r_L^\prime$ satisfies
\begin{equation}
    r_L^{\prime\ \dagger} = r_R \ ,
\label{eq:reflectionless}
\end{equation}
\noindent the left reflection matrix of the full system $r = r_L + t_L^\prime r_L^{\prime \dagger} t_L^{\dagger {-1}}$ vanishes entirely, $r=0$. At the same time, the right reflection matrix vanishes also, $r^\prime =0$, and transmission becomes perfect from both sides: $t^\dagger t = \id$ and $t^{\prime \dagger} t^\prime = \id$. We have therefore established a simple relation to realise a fully transmitting scattering system for which all incident wavefronts $\psi_\mathrm{in}$ are fully transmitted, $\| t\, \psi_\mathrm{in} \|^2 = 1$. From a conceptual point of view, Eq.~(\ref{eq:reflectionless}) constitutes a matrix generalisation of the condition for critical coupling, which is well-known from one-dimensional scattering problems \cite{Baranov2017}. \new{We emphasise here that Eq.~(\ref{eq:reflectionless}) relies on both the amplitude and phase-information of waves to ensure that they interfere appropriately.} Most importantly, our approach does not require knowledge of the microstructure of the disorder in the right medium but only its left-sided (unprimed) reflection matrix $r_{R}$ is necessary to design a fully translucent scattering system. 

\begin{figure*}
\includegraphics[width=\textwidth]{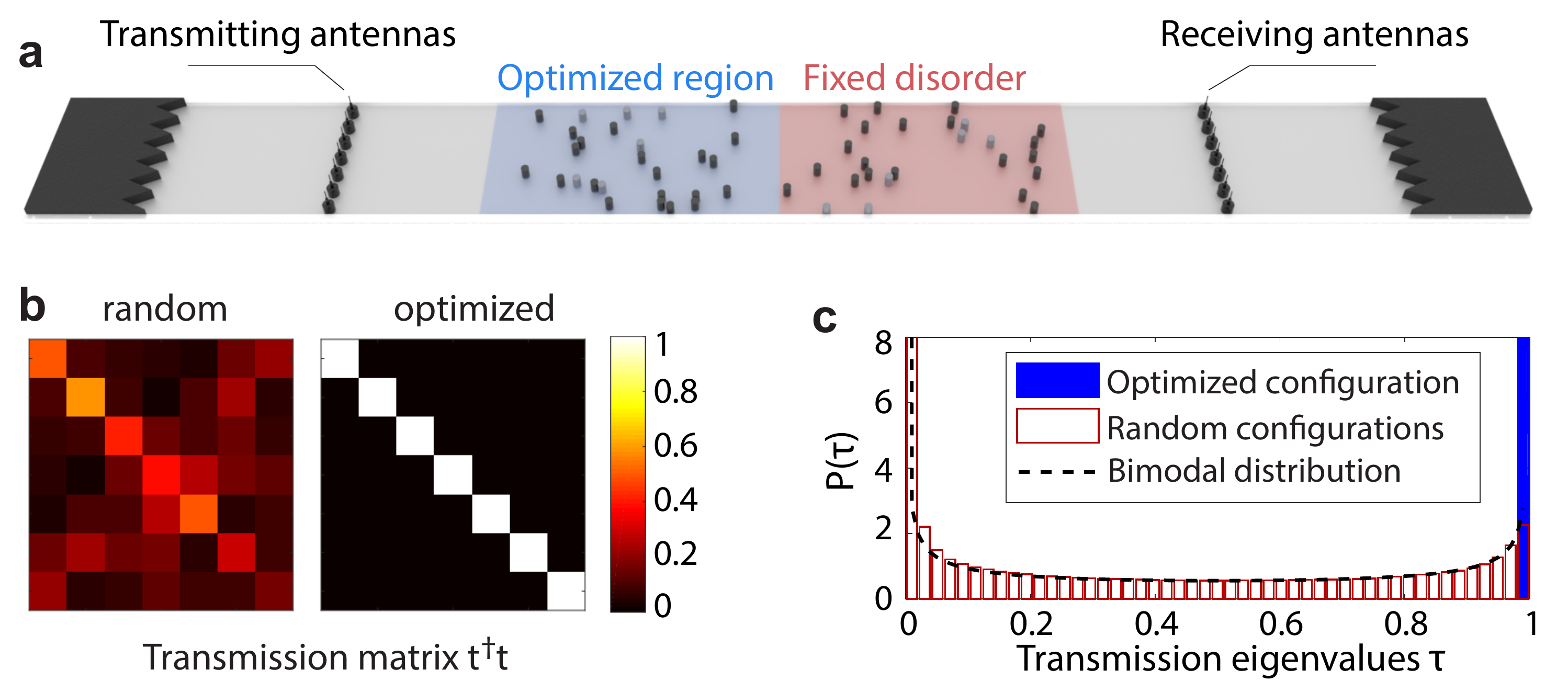}
 \caption{\label{fig:setup} \textbf{Fully transmitting waveguide with disorder.} \textbf{a}, Experimental setup: measurements are carried out with an electromagnetic waveguide of length $L_w = 1$~m, width $W=0.1$~m and height $h=8$~mm. Absorbing foams are placed at each waveguide end to mimic open boundary conditions. We first measure the transmission coefficients between two arrays of 7 wire antennas that are weakly penetrating (3~mm) into the waveguide. The transmission matrix $t(\nu)$ in the basis of the $N$ waveguide modes is then reconstructed by means of a sine transformation (see supplementary material). \textbf{b}, Simulation results for the amplitude of the elements of $t^\dagger t$ for the situation when the complementary medium on the left is randomly chosen (left panel) and optimised for full transmission (right panel). \textbf{c}, The bimodal distribution of transmission eigenvalues $\tau_n$ of $t^\dagger t$ for 2500 random configurations becomes a Dirac distribution $P(\tau) = \delta(\tau-1)$ for the optimised complementary medium as $t^\dagger t$ becomes the identity matrix.}
\end{figure*}

We now demonstrate this concept in practice using a waveguide with internal disorder supporting \new{$N=4$ or $N=7$ transverse modes and a single mode in its vertical dimension over the operating frequency ranges (6.6–7.4~GHz and 10.7-11.7~GHz)}. The propagation of waves through this system is modeled with a two-dimensional scalar Helmholtz equation $\left[ \Delta + n^2(\mathbf{r}) k_0^2 \right] \psi (\mathbf{r}) = 0$. The initial disorder configuration of the given medium consists of a random arrangement of 17 Teflon cylinders and 3 metallic cylinders in a region of length $0.2$~m. Its mean transmission $T = \Sigma_{b,a = 1}^N |t_{ba}|^2 / N$ is equal to $T =  0.64$ at 7~GHz giving an estimated mean free path $\ell_\mathrm{tr} \approx 0.23$~m. \new{At 11.2~GHz we have $T = 0.58$ and $\ell_\mathrm{tr} \approx 0.18$~m.}
%We estimate that the transport mean free path is $\ell_\mathrm{Tr} \approx 0.23m$ giving a mean transmission $T =  0.64$. 

\begin{figure*}
    \includegraphics[width=\textwidth]{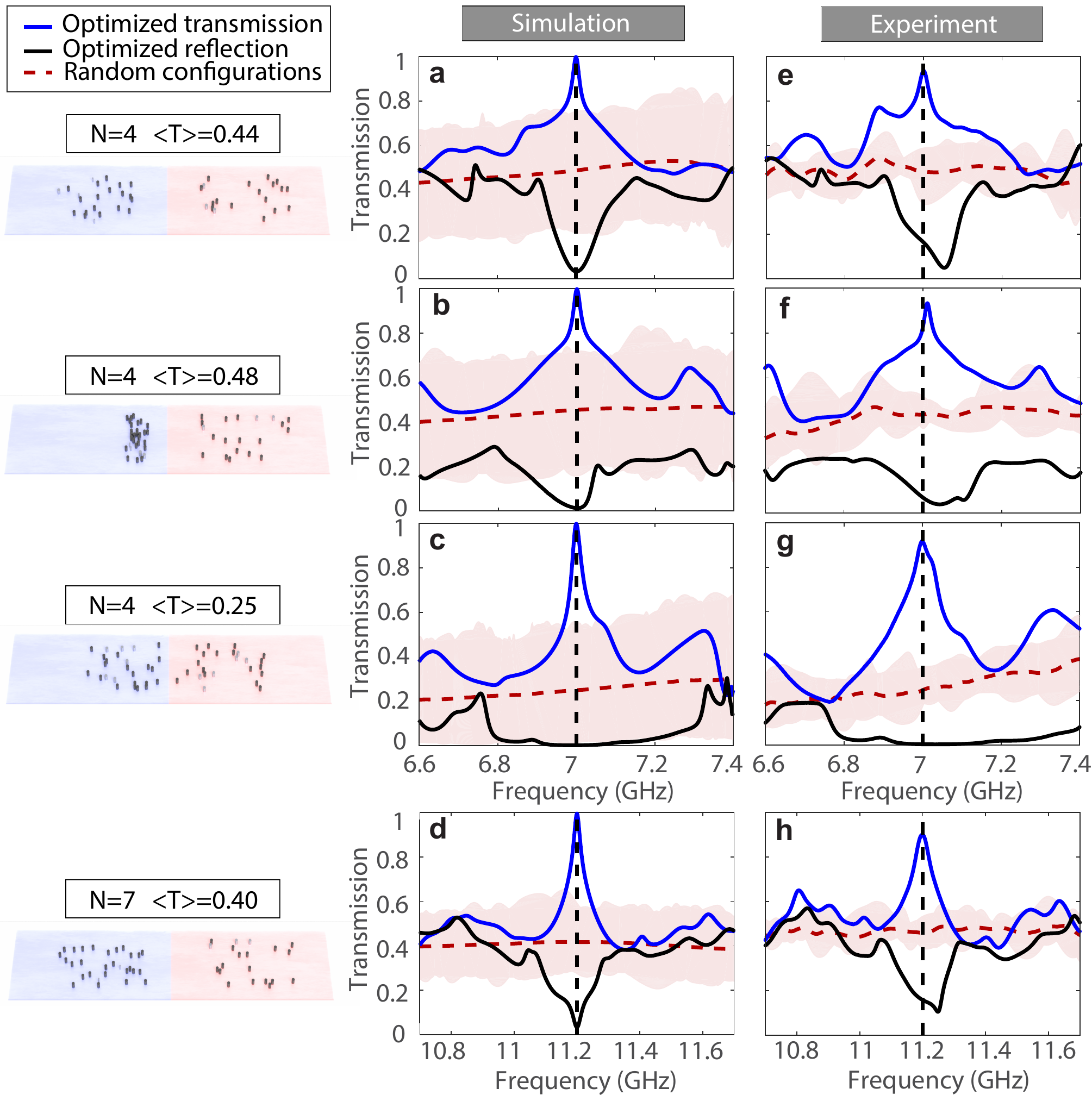}
    \caption{\label{fig:experimental_result} \new{\textbf{Numerical and experimental results.} \textbf{a}-\textbf{d}, Simulated transmission spectra, $T(\nu) = [\Sigma_{n=1}^N \tau_n(\nu)] /N$, for anti-reflection structures that are optimised for maximal transmission (blue line) and maximal reflection (black line) at $\nu_0 = 7$~GHz (\textbf{a}-\textbf{c}) or $\nu_0 = 11.2$~GHz (\textbf{d}). The shaded pink area indicates the range of values obtained for 2500 random configurations with the average being represented by the red dotted line. The fixed disorder consists either of 3 metallic and 17 Teflon cylinders (\textbf{a}, \textbf{b}, \textbf{d}) or 6 metallic and 20 Teflon cylinders (\textbf{c}), while the size of the optimisation region varies to show the design process' flexibility. \textbf{e}-\textbf{h}, Corresponding experimental results. The pink shaded area and the corresponding average values are estimated here from 10 random configurations. \textbf{Left column}, Sketches of the fixed (red) and optimised disorders (blue) associated to each transmission spectrum.}}
\end{figure*}

\section*{Inverse design process}

The critical remaining task is to design a complementary medium with a desired reflection matrix at the design frequency (which we choose to be $\nu_0=7$~GHz \new{or $\nu_0=11.2$~GHz}). Equation (\ref{eq:reflectionless}) requires that the individual values of the average transmission through the complementary medium and the fixed disorder are equal $T_L = T_R$, such that we start the inverse design with a complementary medium that features \new{a combination of Teflon and metallic scatterers (randomly chosen) with a similar reflection to the fixed disorder}. At this initial step, the matrix $t^\dagger t$, \new{featuring the global transmission $t$,} is also random as shown in Fig.~\ref{fig:setup}. The distribution of transmission eigenvalues $\tau$ of $t^\dagger t$ follows a bimodal law for random configurations \cite{Dorokhov1984,Beenakker1997}, \new{$P(\tau) \propto 1/[\tau \sqrt{1-\tau}] $} as expected in scattering media (see Fig.~\ref{fig:setup}c), with an average transmission $\langle \tau \rangle = 0.40$ averaged both over frequencies and configurations. 

In a next step, we gradually optimise the positions of the scatterers in the complementary medium to satisfy the desired matching condition. Since, however, the design space for finding an optimal solution for this inverse problem is enormous, it is insufficient to just work with a random search algorithm. Instead, we need an iterative procedure based on the gradient of the objective (the average transmission of the sample) with respect to the system parameters (the scatterers' positions) \cite{Molesky2018}. Procedures of this kind \cite{Resisi2020,Dinsdale2021} have been implemented in various computational techniques  \cite{Jensen2011,Molesky2018,So2020}, such as for the design of compact and efficient devices in nanophotonics \cite{Liu2011,Shen2015%,Piggott2015%,Yu2017
}, light confinement in strongly scattering disordered media \cite{Riboli2014}, or for analog computing using metastructures \cite{Engheta2019}. 
%Here, we present a novel and powerful optimisation approach to leverage the spatial degrees of freedom of the multiple scatterers in the complementary medium to make its one-sided reflection matrix satisfy the desired matching condition. 
% for engineering the transmission and reflection matrices of a disordered material at will . A similar approach found in microwave technology to engineer the TM relies on adaptively tuning the boundary conditions with reconfigurable metasurfaces \cite{delHougne2019,Frazier2020}. Coherent control of wave propagation can then be implemented with desired arbitrary wavefronts, removing the vexing need for wavefront shaping with spatial light modulator in optics or bulky IQ modulators in microwaves \cite{Jang2018,Liu2020,delHougne2020}. 
%The numerical results obtained are then implemented in a microwave experiment, demonstrating the practical feasibility of our concept in terms of drastic enhancement in the transmission through a disordered environment.
Here we introduce a tailor-made approach to calculate this gradient based on the generalised Wigner-Smith (GWS) operator that has recently been exploited for optimal focusing, micro-manipulation and information retrieval in disordered systems \cite{ambichl_focusing_2017, horodynski_optimal_2020,bouchet_maximum_2021,delHougne2020}. The constituting equation of the GWS operator, $Q_{\mathbf{r}_n}$, reads
\begin{equation}
    \Braket{u | Q_{\mathbf{r}_n} | u} = \Braket{ u | -\mathrm{i} S^{-1} \frac{\mathrm{d}S}{\mathrm{d}\mathbf{r}_n} | u} \propto \int_{0}^{2\pi} \left(\begin{array}{cc} \cos\varphi \\ \sin\varphi \end{array}\right) |\psi_u\left(\mathbf{r}_n\right)|^2 \mathrm{d}\varphi \ , %= \frac{k_0^2}{2}\Braket{\psi_u | \frac{\mathrm{d}n^2(\mathbf{r})}{\mathrm{d}\mathbf{r}_n}|\psi_u}%\mathbf{F}\left( \ket{u}\right),
\end{equation}
where $\mathbf{r}_n$ is the position of the $n$th scatterer, $\psi_u$ the wavefunction resulting from an injection described by a vector of modal amplitudes $\ket{u}$ and the integral traces out the boundary of the scatterer. Given now a certain scattering wavefunction in the near-field of each scatterer, $\psi_u(\mathbf{r}_n)\,$, we can infer which shift of a particular scatterer moves the scattering matrix in the desired direction. The advantage of this approach for calculating the gradient is that the number of electromagnetic simulations we need to perform scales linearly with the number of modes and is independent of the number of system parameters. \new{The scaling of each simulation, however, depends on the electromagnetic solver used.} A more detailed look reveals that for each incoming mode two simulations are required, making our approach exactly as efficient as the state of the art adjoint derivative \cite{Molesky2018}.

\section*{Experimental results}

At the end of the optimisation, the transmission $T$ is extremely close to unity (\new{between $T_{\mathrm{sim}} = 0.995$ and $T_{\mathrm{sim}} = 0.999$ depending on the sample}) such that the matrix $t(\nu_0)^\dagger t(\nu_0)$ practically is the identity matrix (see Fig.~\ref{fig:setup}b, right panel). In this translucent scattering medium, all transmission eigenchannels are naturally open, $\tau_n = 1 \; \forall n$, as seen in Fig.~\ref{fig:setup}c. %The transmission $T$ is enhanced by a factor 2.2 relative to its average for random configurations $\langle T \rangle$ and 
\new{For these samples, the length of the complementary medium is the same as the one of the fixed disorder. However, because only Eq.~(\ref{eq:reflectionless}) needs to be satisfied, we can in principle optimise the complementary medium in a much smaller area. For the same scattering strength, we indeed achieve perfect transmission at 7 GHz using a length as small as a single wavelength. Finally, we repeat the same procedure for an initial sample with stronger disorder (20 Teflon cylinders and 6 metallic cylinders) giving $\ell_\mathrm{tr} \approx 0.095m$ at 7 GHz.} 

The spectra of $T(\nu)$ shown in Fig.~\ref{fig:experimental_result}\new{a-d} all display a pronounced resonance at the design frequency $\nu = \nu_0$. We emphasise that the optimised design of the complementary medium is indeed a rare event as the transmission does not exceed 0.75 in all of the 2500 random configurations we have sampled in our numerics. We also note that the geometry resulting from our optimisation is not a hyperuniform structure \cite{leseur_high-density_2016} nor does it rely on a special mirror symmetry \cite{cheron_broadband-enhanced_2019} as in previous attempts at realising a fully transparent disorder or at enhancing transmission through a barrier; in this sense our approach is broadly applicable (see supplementary material for more information).

In the waveguide experiment, we implement the numerically obtained solution by placing the scatterers at the calculated positions (see Fig.~\ref{fig:setup}a) and by measuring the transmission of microwaves using a set of antennas (see Supplementary Material). The experimental results obtained in this way  nicely reproduce the simulation results with a maximum transmission \new{$T_\mathrm{exp}$ between 0.91 and 0.94 at 7 GHz and $T_\mathrm{exp} = 0.9$ at 11.2 GHz}, see Fig.~\ref{fig:experimental_result}e-h. %The same procedure is then repeated for an initial sample with stronger disorder (20 Teflon cylinders and 6 metallic cylinders) giving $\ell_\mathrm{tr} \approx 0.095m$ at 7 GHz. For an unoptimised complementary disorder, the average transmission is now $\langle T \rangle \approx 0.28$ at 7 GHz. After the optimisation process ($T_{\mathrm{sim}} = 0.999$), the maximum transmission found in measurements is here $T_{\mathrm{exp}} = 0.9$. 
The reduction of $T_{\mathrm{exp}}$ as compared to the numerical values is due to the impact of dissipative losses in the waveguide walls and in the scatterers, which increases with the dwell time of transmitted waves and hence with the disorder strength.
%For completeness, we note that for the two optimised disorders, the peaks of $T(\nu)$ are slightly shifted off by 30~MHz (corresponding to 0.5\% of the central frequency). We attribute this shift to unavoidable manufacturing errors on the positions and the heights of the cylindrical scatterers. %The high sensitivity on the position of the scatterer is a signature of the enhanced the dwell time at the resonant frequency (see the discussion on the dwell time below).
  
\begin{figure*}
    \centering
    \includegraphics[width=\textwidth]{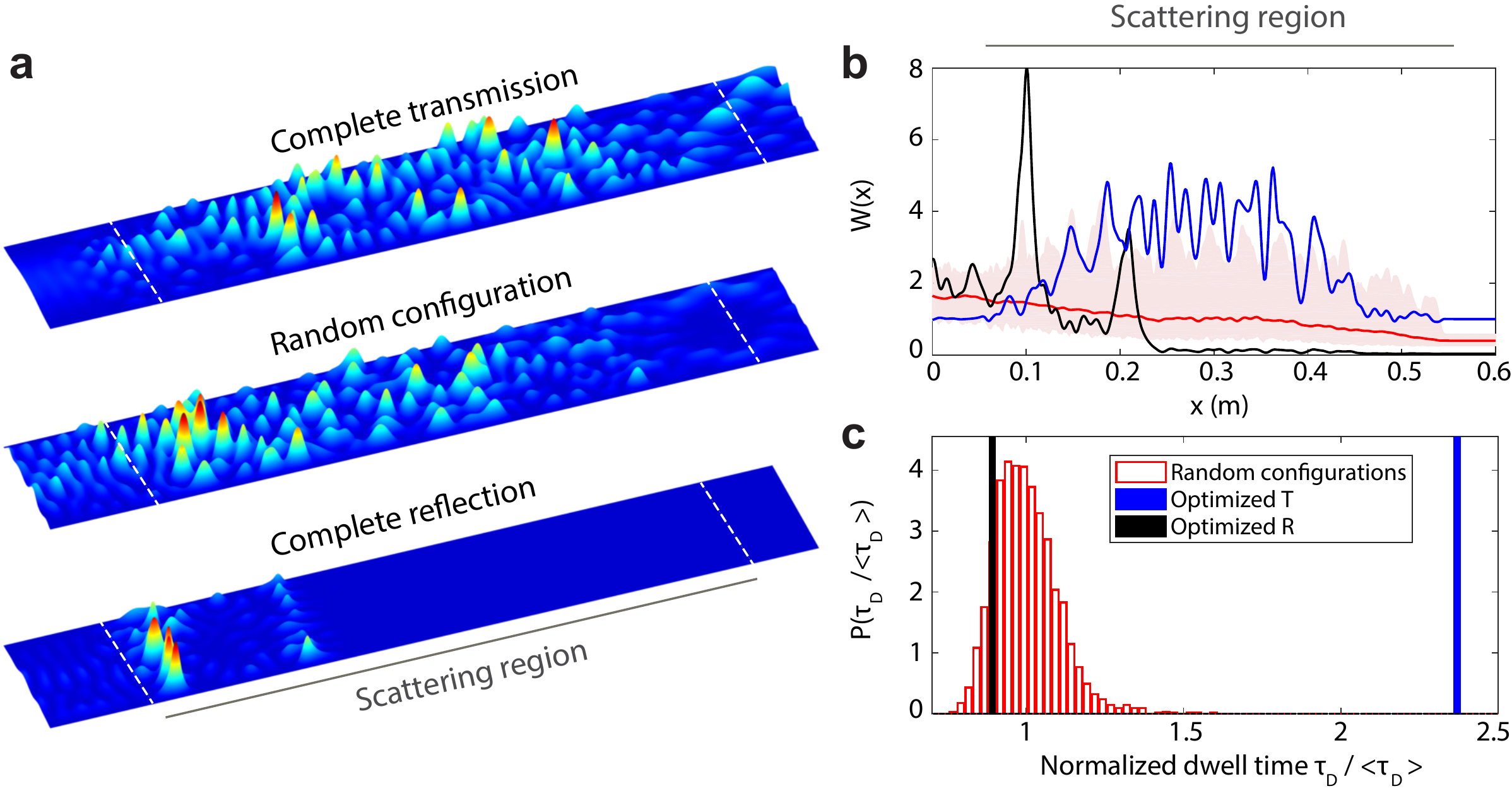}
    \caption{\label{fig:dwelltime} \textbf{Energy stored within the sample.} \textbf{a}, Intensity distribution when injecting the fundamental waveguide mode for a fully transmitting sample (top), for a single random configuration (middle) and for a fully reflecting sample (bottom). The configurations correspond to those of Fig.~3d. The dashed white lines delineate the scattering region. \textbf{b}, Energy profile normalised over the cross-section of the waveguide $W(x)$ for complete transmission (blue line), for complete reflection (black line), and for an averaging over 2500 random configurations (red line) with the entire range of values indicated by the shaded area. \textbf{c}, Probability distribution of the normalised dwell time $\tau_D / \langle \tau_D \rangle$ for random configurations (red bars). In comparison, the dwell time in the fully transmitting sample is enhanced by a factor 2.37 relative to its average over random configuration and reduced by a factor 0.89 in the fully reflecting sample.}
\end{figure*}

To demonstrate the flexibility of our inverse design procedure we use it not only to maximise, but also to minimise transmission and hence maximise reflection through the system with the same number of scatterers. \new{The minimal transmission found numerically (experimentally) now varies between 0.001 and 0.03 (between 0.002 and 0.049) for the samples with $N=4$ and is as small as $T_\mathrm{sim} = 0.03$ ($T_\mathrm{exp} = 0.1$) with $N=7$.}
%The transmissions found numerically (experimentally) \new{now vary between 0.001 and 0.03 (between 0.002 and 0.049) at 7 GHz }is now as small as $T_\mathrm{sim} = 0.04$ ($T_\mathrm{exp} = 0.12 $) for the first disorder and $T_\mathrm{sim} = 0.001$ ($T_\mathrm{exp} = 0.004$) for the disorder with the shorter mean free path. 
Interestingly, we observe that the transmission in these opaque samples \new{can be} reduced over a broad frequency range. This observation is explained by considering that the incident waves penetrate here only weakly into the disordered sample (see Fig.~4) such that no sharply resonant states are excited whose limited bandwidth would make this a narrow-band effect \cite{Shi2015}. 

\section*{Enhanced energy storage}

A remarkable consequence of our medium design concerns the spatial distribution of the intensity $W(x,y)$ within the medium (see Fig.~\ref{fig:dwelltime}a) and with it the dwell time $\tau_D \propto \int dx dy\, W(x,y)$. As has been shown in previous works \cite{Davy2015,Davy2015NCOM,Sarma2016,Durand2019}, open transmission channels in generic disordered media are associated both with an enhanced penetration depth of incoming radiation and an increased dwell time in the disorder. Now that we have created a system that features only open transmission eigenchannels, the intriguing question arising is whether our design naturally also increases both of these quantities for all incoming radiation fields. To verify this explicitly, we show in Fig.~\ref{fig:dwelltime}a the spatial distribution of the intensity and in Fig.~\ref{fig:dwelltime}b the energy density averaged over the cross-section of the waveguide, $W(x) = \langle W(x,y) \rangle_y$, for a translucent sample, for an unoptimised random disorder and for a fully reflecting sample. The peak of maximum energy density for a translucent medium shifts, on average, towards the middle of the sample, in agreement with the profile of \textit{open} channels in random configurations \cite{Davy2015,Davy2015NCOM,Sarma2016,Shi2018}. For samples with thickness $L \gg \ell$, we can therefore estimate that the dwell time is on average increased by a factor $\pi L/(12 \ell)$ relative to its value in an unoptimised random configuration \cite{Davy2015,Durand2019}, for which $\langle W(x) \rangle$ is governed by the diffusion equation and decreases linearly with $x$. The dwell time in translucent samples indeed falls in the tail of the distribution $P(\tau_D)$ computed for many random configurations, see Fig.~\ref{fig:dwelltime}c. In contrast,  for a highly reflecting sample, $W(x)$ approaches the profile of closed channels and decays almost exponentially with $x$, corresponding to a dwell time shorter than $\langle \tau_D \rangle$.

\section*{Conclusion}

\new{Our approach is general and can be broadly applied to other complex systems. In the supplemental material, we extend our inverse design of perfectly transmitting structures to a multichannel cavity coupling four incoming and four outgoing ports. By optimising the position of fifteen metallic scatterers within the cavity, we achieve an average transmission as large as 0.998 numerically and 0.9 experimentally at the selected frequency.} Our results also provide an interesting new perspective for research on reflectionless scattering modes \cite{Dhia2018,Sweeney2020} and coherent perfect absorption \cite{Baranov2017}. While these coherent wave effects have recently been realised also experimentally for disordered systems, all of these studies remained restricted to single specific wavefronts whose absence of reflection or perfect absorption were demonstrated \cite{Pichler2019,Chen2020,delHougne2020CPA}. Our results point the way how to conveniently extend these concepts to multiple incoming wavefronts in parallel. Practically speaking, for the case of a broadband absorber being placed behind a disordered medium, the anti-reflection coating we introduce here would have the interesting effect that the opaqueness of the disorder (appearing as white in the visible spectrum) would get a spectral dip (appearing as black) at the design frequency, at which any incoming wave front penetrates the disorder and gets absorbed perfectly. 

In conclusion, we have demonstrated theoretically and experimentally that a disordered medium can be made fully transmitting to all incoming wavefronts by placing an optimised complementary medium in front of it. %, with a proof of concept using disordered electromagnetic waveguides. 
Since only the reflection matrix of this complementary medium must be engineered, we envision that thin metasurfaces could be used for this purpose, enabling the creation of tailor-made and potentially time-adaptive anti-reflection structures with %The analogy with the concept of CPA also suggests the possibility of creating systems that are made perfectly absorbing to any incoming wavefront. 
%More generally speaking, we expect our results to open up 
fascinating properties for applications in the fields of wireless communications, filtering, energy harvesting and imaging. \new{In the long term, we expect that advances both in computing power and in microfabrication will make our approach applicable to systems with an increasingly large number of modes.}

\bibliographystyle{naturemag}
%\bibliography{bibliography.bib, Bibliography_Michael.bib}% Produces the bibliography via BibTeX.

\section*{Methods}

\subsection*{Numerical modeling and optimisation algorithm}
The experimental set-up is modeled using the 2D scalar Helmholtz equation $[\Delta + k_0^2\varepsilon(\mathbf{r})]\psi(\mathbf{r})=0$, where $\Delta$ denotes the Laplacian in two dimensions, $\varepsilon(\mathbf{r})$ is the spatially varying dielectric constant, $k$ the wave vector and $\psi$ the $z$–component of the electrical field. We solve the Helmholtz equation using an open-source finite element-library (NGSolve) \citeMethods{schoberl_netgen_1997,schoberl_c11_2014}. This allows us also to numerically evaluate the objective function we maximize, which reads 
\begin{equation}
    f = \tr t^\dagger t /N = \tr \left( \id - r^\dagger r\right)/N,\;
    \mathrm{with}\quad r = r_L + t_L^\prime r_R (\id - r_L^\prime r_R)^{-1} t_L\,,\label{eqm:eq5}
\end{equation}
where $t$ ($r$) is the transmission (reflection) matrix of the whole system for a wave injected from the left lead and $N$ is the number of transverse electric modes. Using the unitarity of $S$ we rewrite $f$ in terms of $r$. This has the considerable advantage that we only need the measurable reflection matrix of the fixed disorder, $r_R$, to compute the objective function. The gradient of this objective with respect to the position of the $n$th scatterer is given by,
\begin{linenomath}
\begin{align}
    \nabla_{\mathbf{r}_n} f &= -2 \real \tr \left( r^\dagger \nabla_{\mathbf{r}_n} r \right)/N \,,\;\mathrm{with} \\
    %\nabla_{\mathbf{r}_n} r = \nabla_{\mathbf{r}_n} r_L + \nabla_{\mathbf{r}_n} t_L^\prime r_R (\id - r_L^\prime r_R)^{-1} t_L + t_L^\prime r_R (\id - r_L^\prime r_R)^{-1} \left[ \nabla_{\mathbf{r}_n} r_L^\prime r_R (\id - r_L^\prime r_R)^{-1} t_L + \nabla_{\mathbf{r}_n} t_L\right] \\
    \nabla_{\mathbf{r}_n} r &= \nabla_{\mathbf{r}_n} r_L + \nabla_{\mathbf{r}_n} t_L^\prime r_R M t_L + t_L^\prime r_R M \left[ \nabla_{\mathbf{r}_n} r_L^\prime r_R M t_L + \nabla_{\mathbf{r}_n} t_L\right] \,,
\end{align}
\end{linenomath}
where $ M = (\id - r_L^\prime r_R)^{-1}$ encodes the multiple reflections between left and right disorder. $r_L^\prime$ is the reflection matrix for a state injected from the right.
Following from the block structure of the $S$ matrix the gradient of each block is given by,
\begin{linenomath}
\begin{equation}
    \nabla_{\mathbf{r}_n} S_L = \left(
    \begin{array}{cc}
         \nabla_{\mathbf{r}_n}r_L& \nabla_{\mathbf{r}_n}t_L^\prime \\
         \nabla_{\mathbf{r}_n}t_L& \nabla_{\mathbf{r}_n}r_L^\prime
    \end{array}
    \right) = \mathrm{i}\left(
    \begin{array}{cc}
         r_L& t_L^\prime \\
         t_L& r_L^\prime
    \end{array}
    \right)\left(
    \begin{array}{cc}
         Q_{\mathbf{r}_n}^{11}& Q_{\mathbf{r}_n}^{12} \\
         Q_{\mathbf{r}_n}^{21}& Q_{\mathbf{r}_n}^{22}
    \end{array}
    \right)\,,
\end{equation}
\end{linenomath}
where the matrices $Q_{\mathbf{r}_n}^{ij}$ are the corresponding blocks of the GWS operator associated to the shift of the $n$th scatterer in direction $\mathbf{r}_n$. We note here that when we use Eq.~(\ref{eqm:eq5}) as our objective function we need to take into account evanescent coupling between left and right disorder such that the total transmission is faithfully computed. This can be either done by avoiding evanescent modes altogether (hence the gap between left and right disorder in some configurations) or by using a scattering matrix which includes evanescent channels \citeMethods{PhysRevA.62.012712}. Alternatively one can directly compute the transmission matrix of the whole system.

The GWS operator was originally introduced for optimal micro-manipulation since its eigenvalues are directly proportional to the force applied on the target scatterer \cite{ambichl_focusing_2017, horodynski_optimal_2020}. This has the interesting consequence that we can compute the GWS operator and therefore the gradient of $f$ using the electric field (with which we calculate the force) at the scatterers without the need to move any single scatterer – a feature which makes the evaluation of the gradient independent of the number of system parameters. To be more precise, we only have to do $2N$ simulations, where $N$ is the number of modes. We access the matrix elements of $Q_{\mathbf{r}_n}$ by
\begin{linenomath}
\begin{align}
\begin{split}
    4\left[Q_{\mathbf{r}_n} \right]_{ij} = &\ \mathbf{F} \Big(\ket{u_i} + \ket{u_j}\Big)  - \mathbf{F}\Big(\ket{u_i} - \ket{u_j}\Big) \\
    &+ \mathrm{i} \mathbf{F}\Big(\ket{u_i} - \mathrm{i}\ket{u_j}\Big) - \mathrm{i} \mathbf{F}\Big(\ket{u_i} + \mathrm{i}\ket{u_j}\Big),
\end{split}
\end{align}
\end{linenomath}
where the $\ket{u_i}$ are a complete and orthonormal set of incoming scattering states (e.g., the different waveguide modes) and $\mathbf{F} \Big(\ket{u}\Big)$ denotes the force transferred by the electric field onto the $n$th scatterer when we inject a wave in the state $\ket{u}$ into the system. This simple reformulation allows us to simulate the electric field inside the scattering region and then calculate $Q_{\mathbf{r}_n}$ for every single scatterer with which we can build the gradient of $f$ without the need to move every single scattere, simulate the system and calculate the gradient. In the case of a dielectric target the force is calculated by,
\begin{linenomath}
\begin{align}
    \mathbf{F} \Big(\ket{u}\Big) = \frac{Rk^2(\varepsilon-1)}{2} \int_{0}^{2\pi} \left(\begin{array}{cc} \cos\varphi \\ \sin\varphi \end{array}\right) |\psi_u\left(\rho=R\right)|^2 \mathrm{d}\varphi\,,
\end{align}
\end{linenomath}
and for a metallic one by,
\begin{linenomath}
\begin{align}
    \mathbf{F} \Big(\ket{u}\Big) = -\frac{R}{2} \int_{0}^{2\pi}  \left(\begin{array}{cc} \cos\varphi \\ \sin\varphi \end{array}\right) \left|\partial_\rho\psi_u\left(\rho=R\right)\right|^2 \mathrm{d}\varphi\,,	
\end{align}
\end{linenomath}
where $\psi_u$ is the electric field distribution inside the scattering region for a wave injected in the state $\ket{u}$.\\\indent
The optimisation is constrained by the strict condition that no two scatterers can overlap. In order to enforce this at every step, we use the following algorithm: while $f<0.999$ (i.e., the transmission is smaller than $99.9\%$), we compute $Q_{\mathbf{r}_n}$ for all scatterers and use it to calculate $f(\mathbf{r}_n + \Delta\mathbf{r}_n)$ with $S(\mathbf{r}_n + \Delta \mathbf{r}_n) = S(\mathbf{r}_n) \exp (\mathrm{i}|\Delta \mathbf{r}_n| Q_{\mathbf{r}_n}) $ for all possible changes in the position of scatterers. We then pick the direction of largest increase in the total transmission, which does not result in an overlap of scatterers. If we cannot find such a change for the given step size, we reduce the step size by a factor of $1.5$. Otherwise we check if $f(\mathbf{r}_n + \Delta\mathbf{r}_n) \geq f(\mathbf{r}_n) + c|\Delta \mathbf{r}_n| \left( \nabla_{\mathbf{r}_n}f\right)^2$, with $c=0.025$ is fulfilled. If this is the case we update the geometry, if not we also reduce the step-size.

\subsection*{Experimental setup}

We first measure the field transmission coefficients $t'(y_2,y_1)$ between 7 transmitting and 7 receiving antennas at locations $y_1$ and $y_2$ respectively. The spacing between each transmitting or receiving antennas is equal to $W/8$, where $W$ is the width of the waveguide. The elements $t_{mn}(\nu)$ of the transmission matrix $t(\nu)$ in the basis of waveguide modes is then reconstructed by means of a two-fold sine transformation:

\begin{align}
\begin{split}
    t_{mn} = \sum_{y_1,y_2} &t'(y_2,y_1) \sqrt{k_n(\nu) k_m(\nu)}\,\text{sin}\left(\frac{m\pi}{W}y_2\right) \text{sin}\left(\frac{n\pi}{W}y_1\right) .
\end{split}
\end{align}

\noindent The transverse mode number $n$ is given by $k_n = (2\pi / c_0) \sqrt{\nu^2 - (n \nu_c)^2}$, the cut-off frequency is $\nu_c = c_0 / 2h$, $c_0$ is the speed of light and $h = 8$~mm is the height of the waveguide. \new{We verify that the transmission matrix in the basis of waveguide modes is diagonal for an empty waveguide (see SM).}

Our antennas have small penetration depths (3~mm) and are thus weakly coupled to the waveguide as a result of an imperfect impedance matching. We therefore normalize the transmission for each waveguide mode $n$, $T_n(\nu) = \Sigma_{m} |t_{mn}(\nu)|^2$, by its value for an empty waveguide, $T^0_n(\nu)$. The mean transmission through a disordered waveguide is then estimated by:

\begin{equation}
    {T}(\nu) = \frac{1}{N} \sum_{n=1}^N \frac{T_n(\nu)}{T_n^0(\nu)} .
\end{equation}

This allows us to compare the experimental results with numerical simulations in Fig.~3 of the main text. \new{To further confirm our normalization procedure, we show in the SM that the distribution of transmission eigenvalues found experimentally for random samples is bimodal in agreement with diffusion theory and numerical results.}

\section*{Data and materials availability}
The data that underlie the plots within this paper and other findings of this study are available from the corresponding authors on reasonable request.

\bibliographystyleMethods{naturemag}
%\bibliographyMethods{bibliography_methods.bib}

\section*{Acknowledgements}
 We acknowledge useful discussions with D. B. Phillips. This publication was supported by the European Union through the European Regional Development Fund (ERDF), by the French region of Brittany and Rennes M{\'e}tropole through the CPER Project SOPHIE/STIC \& Ondes, and by the Austrian Science Fund (FWF) through project P32300 (WAVELAND). C.F. acknowledges funding  from the French “Minist{\`e}re de la D{\'e}fense, Direction G{\'e}n{\'e}rale de l’Armement”. M.D. acknowledges the Institut Universitaire de France. The computational results presented were achieved using the Vienna Scientific Cluster (VSC).
\section*{Author contributions} M.D. proposed the project. Numerical simulations were carried out by M.H and M.K. under the supervision of S.R. Measurements and data evaluation were carried out by C.F. and M.D. M.H., S.R. and M.D. wrote the manuscript with input from all authors.
\section*{Competing interests} The authors declare no competing interests.

\section*{Supplementary materials}
Supplementary information is available for this paper.

\section*{Corresponding authors}
stefan.rotter@tuwien.ac.at\\
matthieu.davy@univ-rennes1.fr

\providecommand{\noopsort}[1]{}\providecommand{\singleletter}[1]{#1}%

\clearpage

\renewcommand{\thefigure}{S\arabic{figure}}
\renewcommand{\theequation}{S\arabic{equation}}
\setcounter{equation}{0}
\setcounter{figure}{0}
\setcounter{section}{0}

\font\myfont=cmr12 at 20pt

%\author{Matthieu Davy}
%\affiliation{Univ.\ Rennes, CNRS, IETR (Institut d'{\'E}lectronique et des Technologies du num{\'e}Rique), UMR–6164, F-35000 Rennes, France}
%\author{Matthias K\"uhmayer}
%\affiliation{ Institute for Theoretical Physics, Vienna University of Technology (TU Wien), A-1040 Vienna, Austria}
%\author{Sylvain Gigan}
%\affiliation{ Laboratoire Kastler Brossel, Universit{\'e} Pierre et Marie Curie, {\'E}cole Normale Sup{\'e}rieure, CNRS, Coll\`{e}ge de France, F-75005 Paris, France}
%\author{Stefan Rotter}
%\affiliation{ Institute for Theoretical Physics, Vienna University of Technology (TU Wien), A-1040 Vienna, Austria}
%\date{\today}

\title{\noindent \LARGE Supplementary Material: Customized anti-reflection structure for perfect transmission through complex media}
%\begin{titlepage}
\maketitle
%\end{titlepage}

\section{Comparison to hyperuniform medium}

\begin{figure}[t!]
    \centering
    \includegraphics[width=\columnwidth]{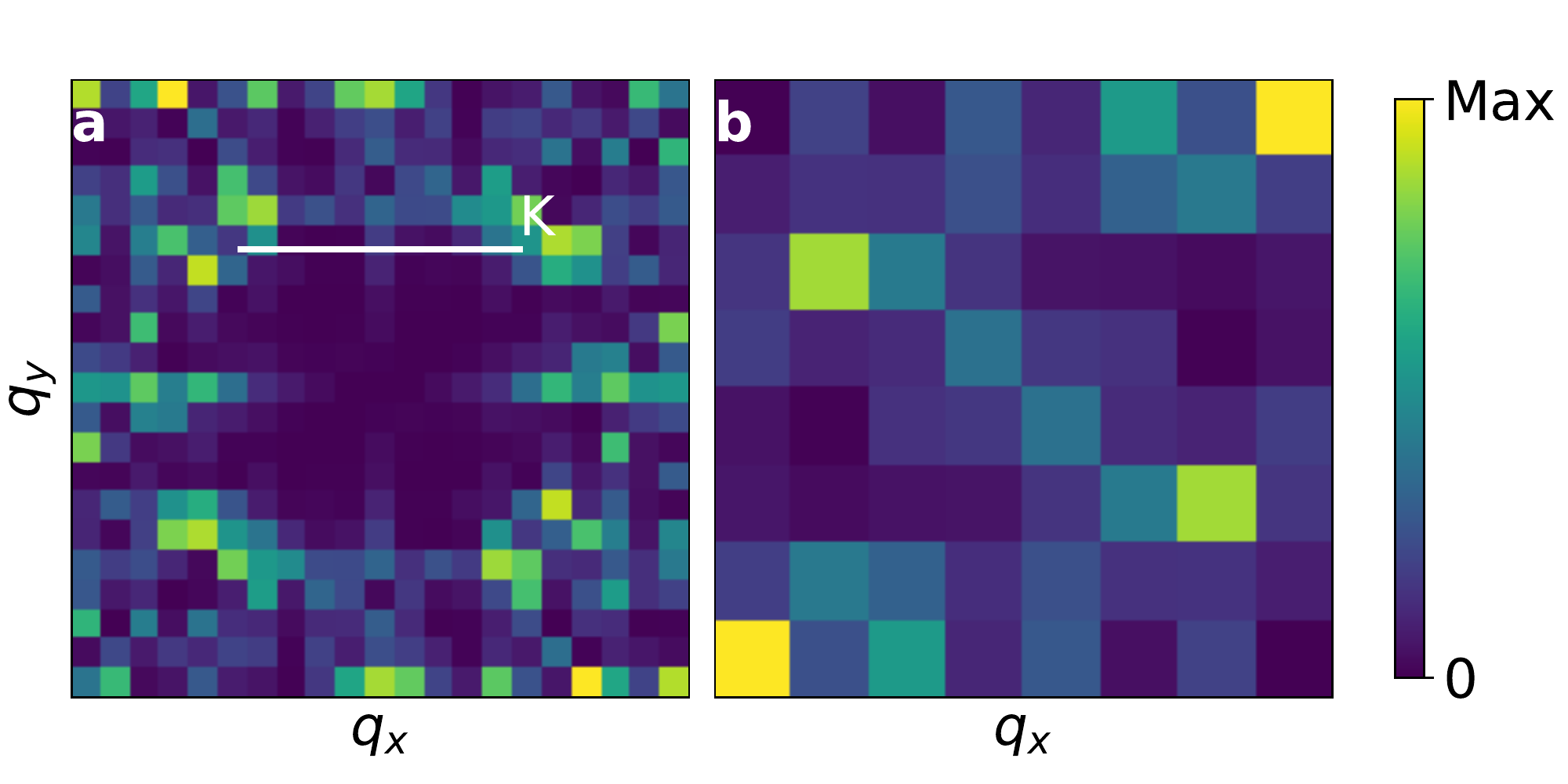}
    \caption{Structure factor (\textbf{a}) for a hyperuniform medium and (\textbf{b}) for a fully translucent disorder resulting from our design protocol. The hyperuniform medium consists of 100 scatterers, while our inverse designed medium features 52 scatterers, corresponding to the stronger variant disorder presented in the main text.}
    \label{fig:figa1}
\end{figure}

Here we investigate if the fully transmitting disorder we designed is hyperuniformly distributed, since hyperuniform media can be transparent, while still being dense enough that transparency is not to be expected \cite{leseur_high-density_2016}. The criterion for a hyperuniform medium to be transparent is that the structure factor $S(\mathbf{q})$, which is defined as
\begin{equation}
    S(\mathbf{q}) = \frac{1}{N} \left|\sum_{j=1}^N \exp(\mathrm{i \mathbf{q}\cdot\mathbf{r}_j}) \right|^2,
\end{equation}
vanishes in a neighbourhood of $|\mathbf{q}|=0$. Here, $N$ is the number of scatterers at positions $\mathbf{r}_j$, while $\mathbf{q}$ denotes the wavevector. In order to find a hyperuniform medium one minimizes a potential
\begin{equation}
    \begin{split}
        \phi(\mathbf{r}_1, ..., \mathbf{r}_N)  = \sum_{j,l=1}^N \frac{\sin\left[ (2P+1)\pi(x_j-x_l)/L\right]}{\sin\left[ \pi (x_j-x_l)/L \right]} \\ \times \frac{\sin\left[ (2P+1)\pi(y_j-y_l)/L\right]}{\sin\left[ \pi (y_j-y_l)/L \right]},
    \end{split}
\end{equation}
where $P=KL/(4\pi)$ and $L$ is the length of the quadratic system. The extent of the region where $S(\mathbf{q})$ vanishes is given by $K$. We show in Fig.~\ref{fig:figa1}a the structure factor of a hyperuniform medium with periodic boundary conditions. It is clearly visible that in a region of size $K$ the structure factor vanishes, resulting in a hyperuniform structure. When we compute, for comparison, the structure factor for the fully transparent medium resulting from our novel design principle, we find that $S(\mathbf{q})$ does not vanish around $|\mathbf{q}|=0$. This shows that the fully translucent disordered medium resulting from our optimisation procedure is not a hyperuniform medium.
\begin{figure}[t!]
    \centering
    \includegraphics[width=\columnwidth]{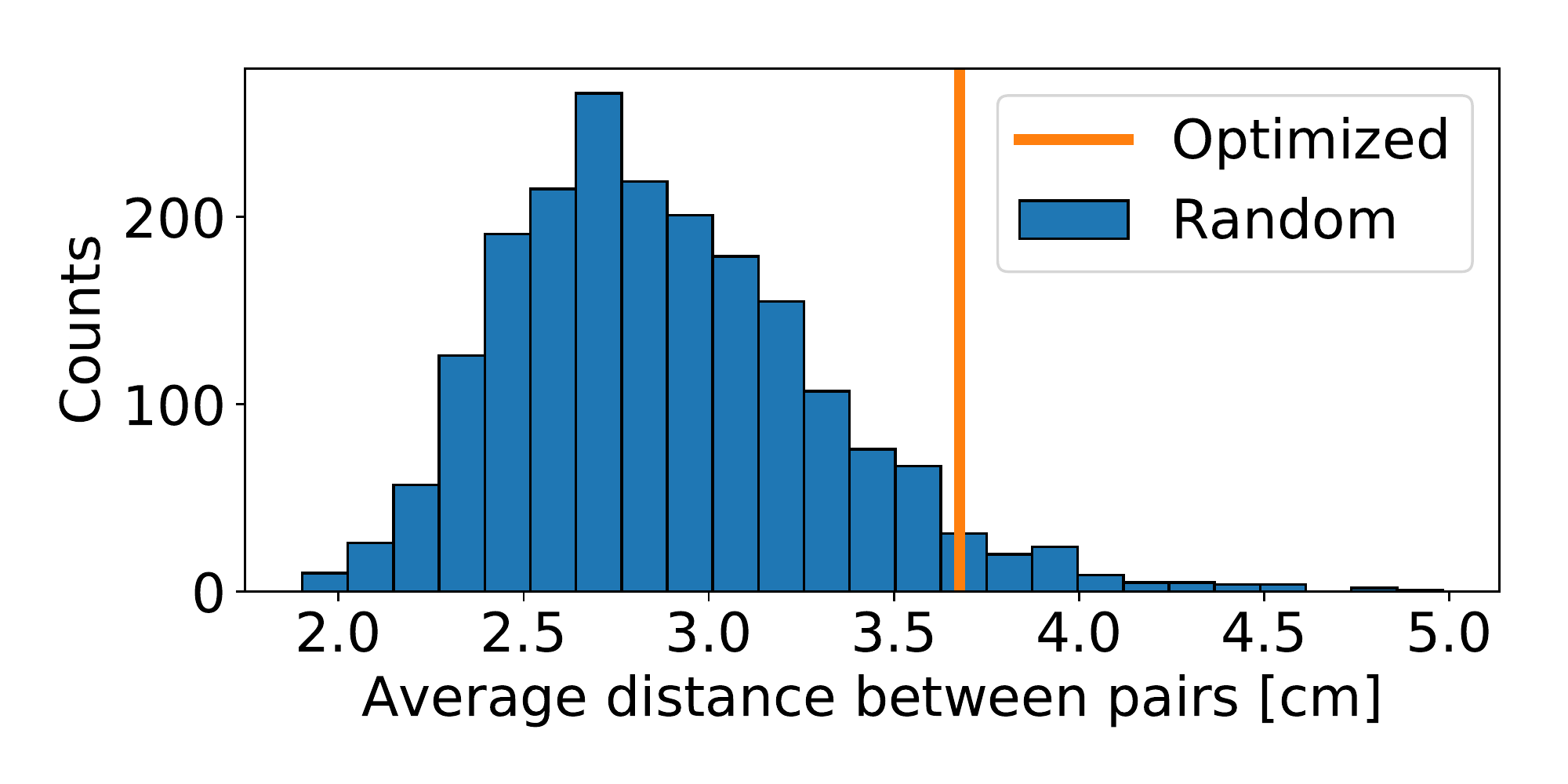}
    \caption{Histogram of the distance of 2000 random disorders to a mirror symmetric disorder (see text). A distance of $0$ would mean that we have perfectly mirror symmetric medium, while larger distances signify that we move away from mirror symmetry. The orange line shows the distance for the optimised medium.}
    \label{fig:figa2}
\end{figure}

\section{Comparison to mirror medium}

It was recently demonstrated that a disordered slab featuring a mirror symmetry along its transverse axis has an increased transmission compared to a fully random medium \cite{cheron_broadband-enhanced_2019,DavyAPL2021}. Here we investigate whether such a symmetry is also (partially) responsible for the full transmission through a disorder. To do so we mirror the fixed disorder we use in Fig.~3b of the main text. We then pair up every scatterer of the mirror disorder with one of an optimised or random disorder of the same material such that the total distance between all pairs is minimal. This distance then tells us how close the disorder is to its mirror-symmetric counterpart. We find (see Fig.~\ref{fig:figa2}) that the optimised disorder is even farther from the mirror disorder than the mean of random configurations (by a factor of about $1.3$).

\section{Probability for randomly sampling a fully transmitting disorder}
\begin{figure}
    \centering
    \includegraphics[width=\columnwidth]{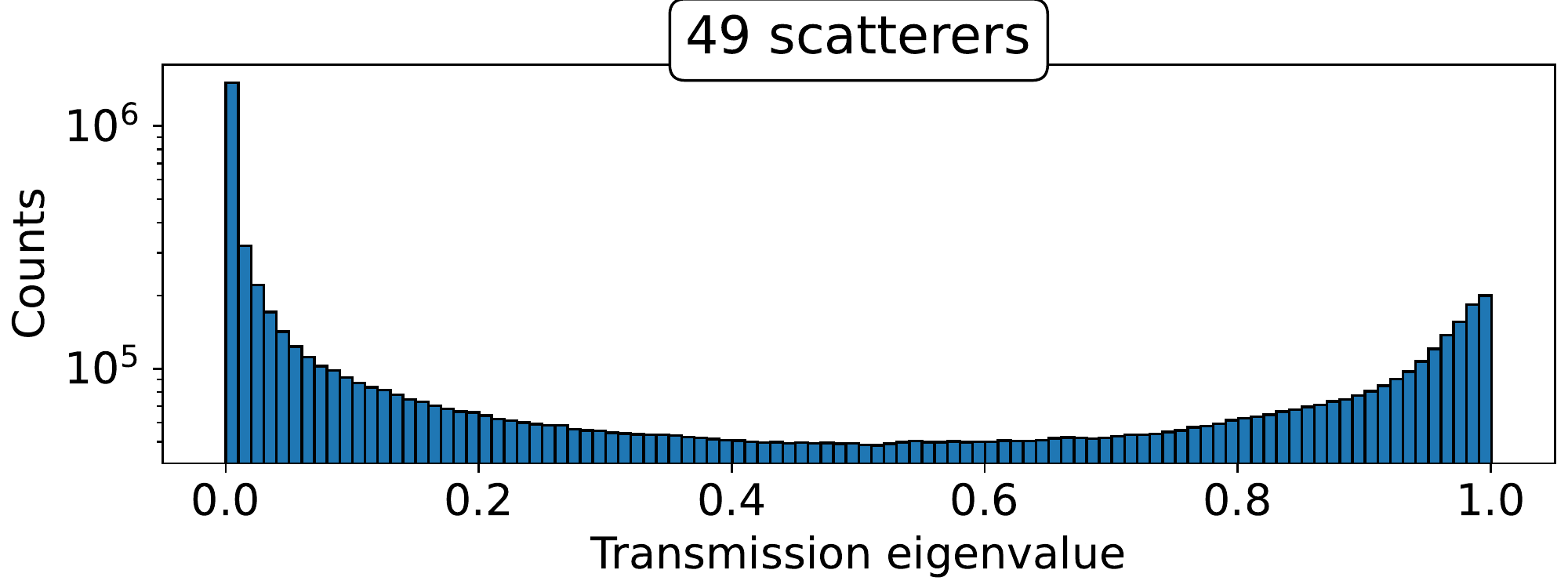}
    \caption{\new{Histogram of the transmission eigenvalues of 2500 random configurations composed of 49  scatterers. Note that we have scanned every sample between 10.7 and 11.7 GHz with a resolution of 501 data points within this frequency window.}}
    \label{fig:figa3}
\end{figure}

\new{An interesting question is whether it would also be possible to find a fully transmitting disorder merely by sampling the positions of scatterers randomly, under the constraint that the disorder overall has the same composition and thickness as the sample considered in Fig.~3\textbf{d} of the main manuscript. A first hint that this is a hard task is the observation that the maximal transmission of random disorders with the same number of scatterers as the one considered in Fig.~3\textbf{d} at 11.2GHz is $T\approx 0.58$. Here we make this observation more quantitative and find that the probability of observing a transmission $T \geq 0.995$ (the one achieved with our inverse designed media), when randomly choosing the position of scatterers, is $6.3\times 10^{-14}$. We arrive at this conclusion by calculating from the bimodal distribution of transmission eigenvalues (see Fig.~\ref{fig:figa3}) a probability of $1.3\times 10^{-2}$ that \textit{one} transmission eigenchannel has transmission $T \geq 0.995$. Since the waveguide we investigate supports 7 propagating modes this probability has to be taken to the power of seven (not taking into account the repulsion of eigenvalues, which will reduce this probability even further). We thus arrive at the conclusion that it is nearly impossible to encounter a fully transmitting disorder by chance.} 

\section{Analysis of the optimisation}
\begin{figure}[t!]
    \centering
    \includegraphics[width=\columnwidth]{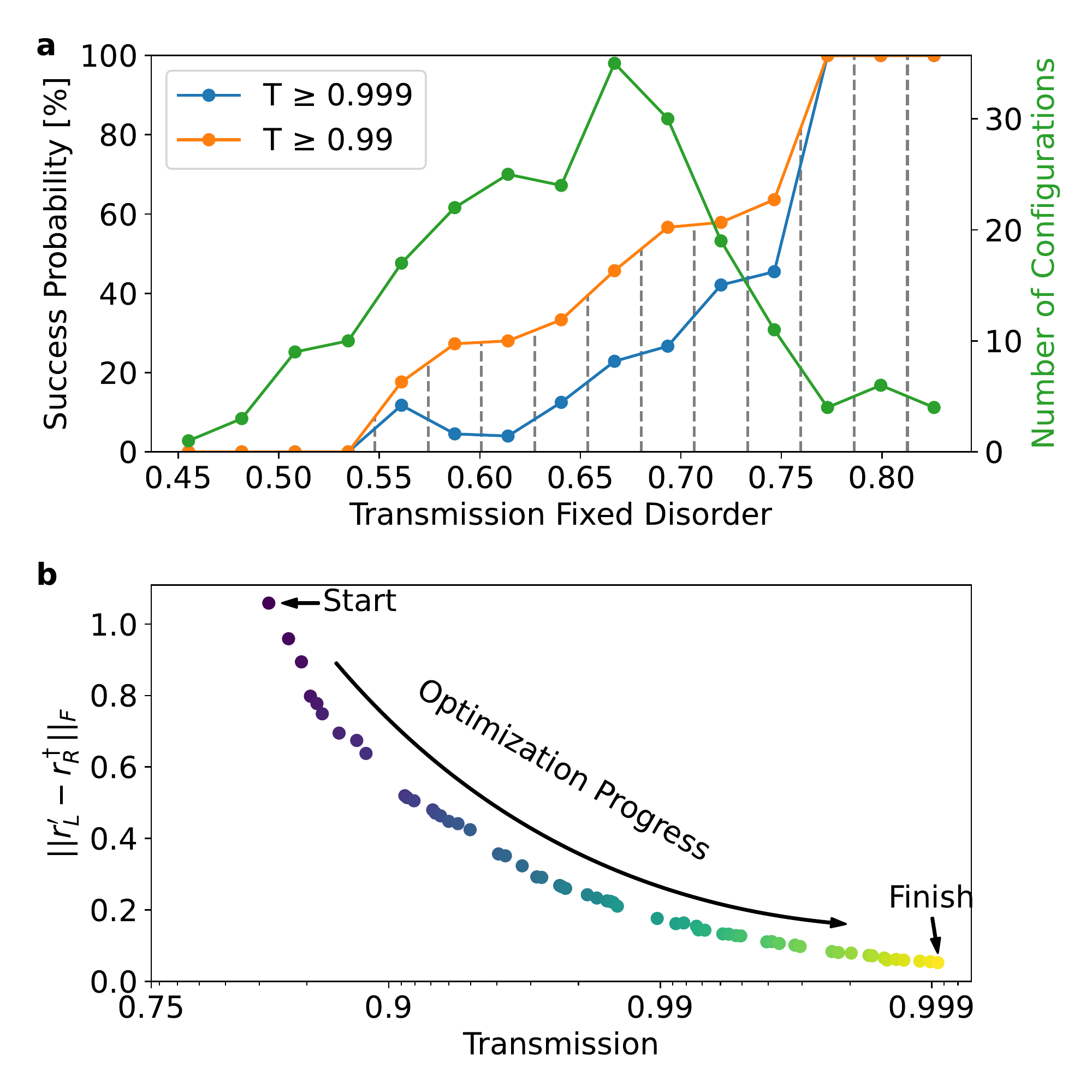}
    \caption{\new{\textbf{a}, Probability of successfully designing a fully transmitting medium as a function of the fixed disorders transmission. Success is defined as a transmission, $T\geq 0.999$ (blue line) or $T\geq 0.99$ (orange line). Note that we are using binned data here with the bin edges indicated by the grey dashed lines. The number of inverse design processes in each bin is given by the green line. The total number of configurations is $220$. \textbf{b}, Plot of the critical coupling condition's Frobenius norm, $||r_L^\prime - r_R^\dagger||_F$, versus the total transmission for each step of the optimisation process. The progress in the inverse design is marked by the colormaps transition from dark blue to bright yellow.}}
    \label{fig:figa6}
\end{figure}

\new{Here we look at the probability to converge to a fully transmitting disorder. To do this we randomly sample fixed disorders consisting of 3 Aluminium and 17 Teflon scatterers. We then choose one random initial configuration for the optimisation region for each fixed disorder and start the inverse design process at 7GHz (4 open modes). The results in dependence of the transmission of the fixed disorder are presented in Fig.~\ref{fig:figa6}\textbf{a}. We see that higher transmission of the fixed disorder leads to a higher probability that the design process converges for one random initial guess. Note that the statistics is not very precise at low transmission due to the low number of configurations.\\\indent
To interpret this result, we note that the gradient descent optimisation used here deterministically follows the path of the steepest gradient until it reaches a local minimum. Since the objective function in the present work is non-convex, different initial conditions lead to different local minima, none of which is guaranteed to be the global minimum. This is in general an unavoidable problem in inverse design \cite{Kuang:20}, but as we demonstrate we are still able to design configurations with a transmission of practically unity. Global methods like genetic algorithms are however no alternative since they are computationally more expensive (by orders of magnitudes) among other problems \cite{Molesky2018}.\\\indent
We also examine one such optimisation process in detail to show that maximising the transmission also leads us to fulfil Eq.~(3) of the main text. In Fig.~\ref{fig:figa6}\textbf{b} we show total transmission and the Frobenius norm of Eq.~(3), $||r_L^\prime - r_R^\dagger||_F$, for each step of one optimisation process shown in Fig.~\ref{fig:figa6}\textbf{a}. While the total transmission monotonically (as required) approaches $T=0.999$ (the stopping condition), the generalised critical coupling condition is fulfilled better and better.}

\section{Steep angles of incidence}
\begin{figure}[t!]
    \centering
    \includegraphics[width=\columnwidth]{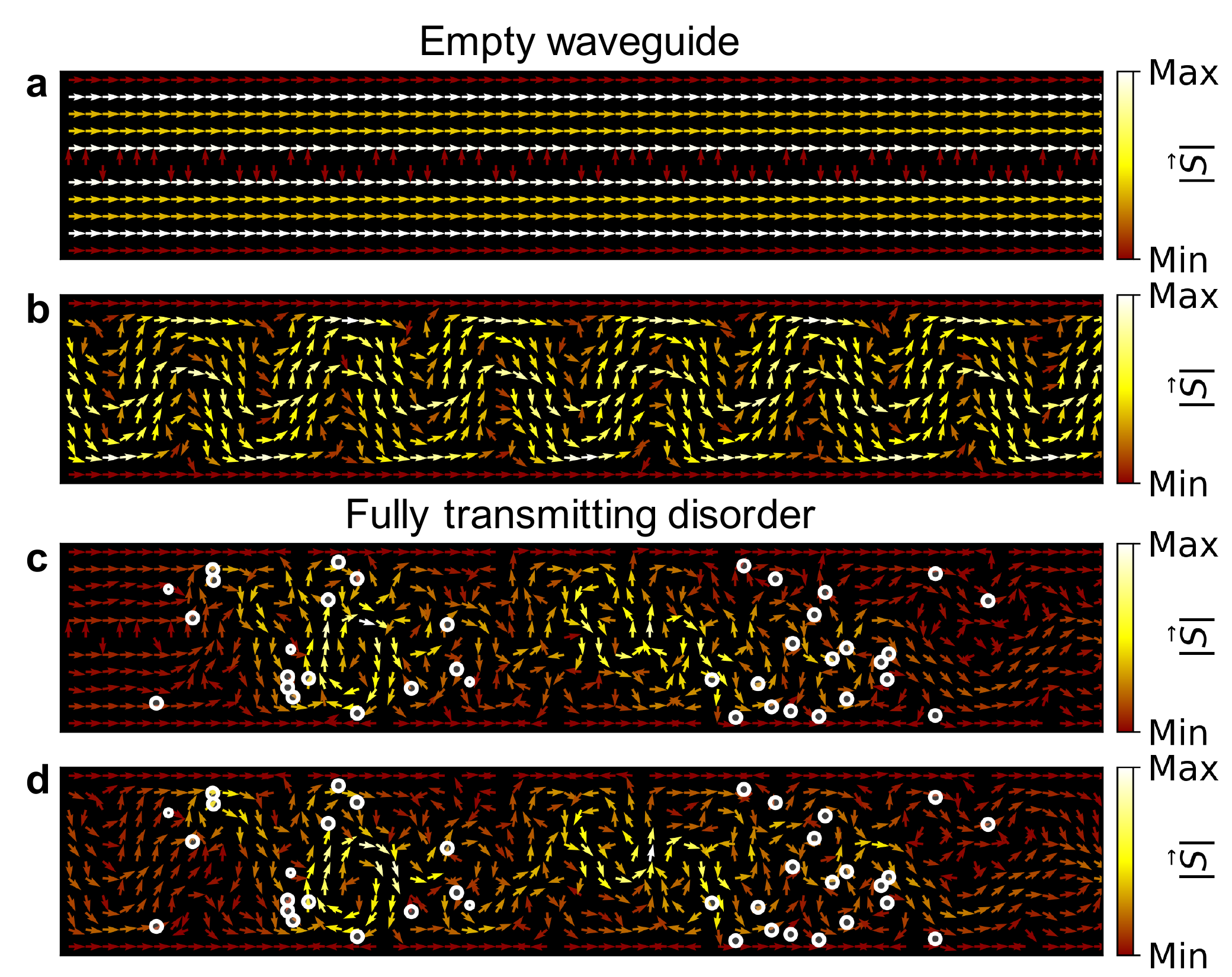}
    \caption{\new{Poynting vector distribution in an empty waveguide (\textbf{a}, \textbf{b}) and a waveguide filled with a fully transmitting disorder (\textbf{c}, \textbf{d}). The incoming wave is either the highest order mode (out of 4) (\textbf{a}, \textbf{c}) or the state optimised for a steep angle of incidence (\textbf{b}, \textbf{d}). The white circles indicate the position of the scatterers.}}
    \label{fig:figa7}
\end{figure}

\new{Here we investigate the role of waveguide modes with very steep angles of incidence. To be more precise, we design a fully transmitting disorder at a frequency of $6.04$ GHz. This means that the $4$th (highest) mode is barely excited resulting in an angle between $\mathbf{k}$ and the longitudinal direction of about $83^{\circ}$. Despite this steep angle of incidence the average transmission of this disorder is still $T=0.998$, while the transmission of the last mode (with the steepest angle) is $T=0.999$.\\\indent 
To further illustrate this result, we show in Fig.~\ref{fig:figa7} the Poynting vector for an empty waveguide and an optimised disorder, when we either inject the highest order mode or a state $\mathbf{u}$ with a high angle of incidence. We construct this state by numerically optimising the ratio of $\mathbf{u}^\dagger k_y \mathbf{u}$ and $\mathbf{u}^\dagger k_x \mathbf{u}$. Here $k_x$ and $k_y$ are the operators that give us for a particular incoming wavestate its $\mathbf{k}$-components. We clearly see in the empty waveguide that we found a state bouncing up and down on the waveguide walls. In the waveguide filled with a perfectly transmitting disorder this state still has perfect transmission since the disorder was designed to be perfectly transmitting for \textit{all} incoming states.}

\section{Effects of absorption}
\begin{figure}
    \centering
    \includegraphics[width=\columnwidth]{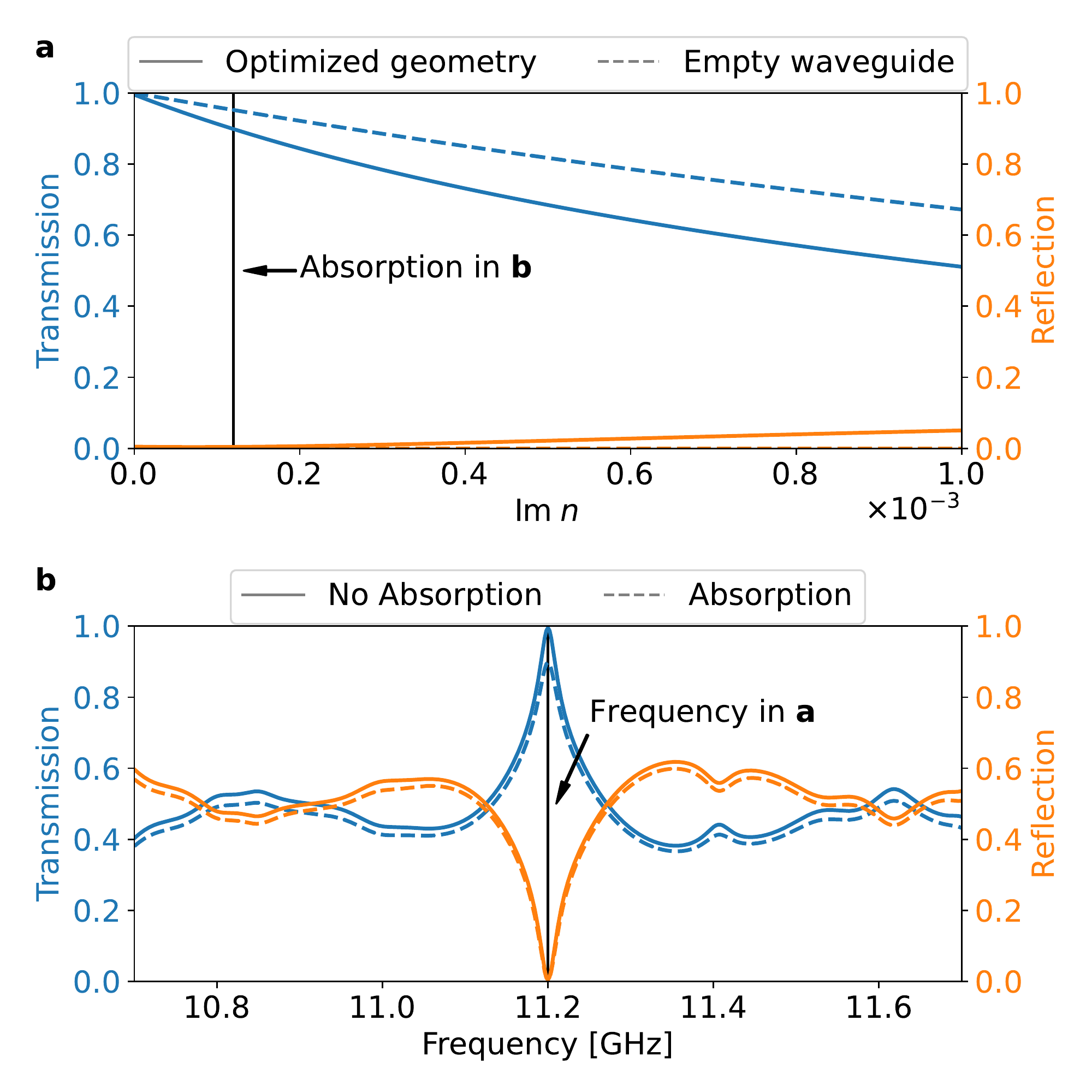}
    \caption{\new{\textbf{a}, Transmission (blue) and reflection (orange) plotted over the imaginary part of the global refractive index for the geometry optimised for 11.2 GHz (solid lines) and an empty waveguide (dashed lines). \textbf{b}, Transmission (blue) and reflection (orange) spectrum of the sample with 49 scatterers. The spectra without absorption are depicted by a solid line, while transmission and reflection curves with added absorption are indicated with dashed lines.}}
    \label{fig:figa4}
\end{figure}

The perfectly transmitting media we present in the main manuscript were designed without considering the effects of absorption. Here we show numerically \new{and analytically} that adding constant global absorption still results in zero reflection across all input channels, while the transmission deviates unavoidably from unity (see Fig.~\ref{fig:figa4}). \new{To be more precise, we first investigate in Fig.~\ref{fig:figa4}\textbf{a} how the transmission and reflection change when we scan the absorption strength in the geometry featuring 49 scatterers designed at 11.2 GHz. We implement the absorption by adding an imaginary part between $n_I=0$ and $n_I=10^{-3}$ to the refractive index everywhere. We find that while the transmission decreases from unity immediately the reflection remains zero for a much wider range of absorption strength. We corroborate this result in Fig.~\ref{fig:figa4}\textbf{b}, where we show a frequency scan of the transmission with and without absorption that features a pronounced dip in the reflection at 11.2 GHz. In this case we use an imaginary part $n_I=1.2\times 10^{-4}$ since it results in a transmission of $T_\mathrm{Abs}\approx0.9$, comparable to the one found in the experiment. Also the reflection at the target frequency (11.2 GHz) is $R_\mathrm{Abs} = 0.0042$, which is comparable to the one found without absorption ($R=0.0049$).\\\indent
We put this empirical observation on more solid grounds by considering the globally uniform imaginary part as a complex shift of the frequency to $\omega + \mathrm{i}\alpha/2$, where $\alpha = 2kn_I$ is the absorption rate \cite{beenakker_distribution_2001}. Under the assumption of small dissipation we can then expand the scattering matrix into, 
\begin{equation}
    S_a(\omega + \mathrm{i}\alpha/2) \approx S(\omega) \left[ \id -\frac{\alpha}{2}Q(\omega)\right],
\end{equation}
where $Q$ is the Wigner-Smith time-delay operator and the subscript $a$ denotes absorption (terms without it are evaluated at zero absorption). Using $t^\dagger t = \id$ and $r=0$ we find that
\begin{align}
    t_a^\dagger t_a^{~} \approx \id - \alpha Q_{11}\quad \mathrm{and}\quad r_a^\dagger r_a^{~} \approx \frac{\alpha^2}{4} \partial_\omega r^\dagger \partial_\omega r.
\end{align}
Here $Q_{11}$ is the time-delay operator for a wave impinging onto the disorder from the left. This shows that the reflection deviates from zero only in $\mathcal{O}(\alpha^2)$, while the transmission decreases from unity already in first order, making the zero-reflection medium robust to absorption.}

\section{Experimental setup}

\new{
The transmission matrix $t'(y_2,y_1)$ is first measured between two arrays of seven pointlike  antennas with small penetration depth. As shown in the main text, we then reconstruct the transmission matrix in the basis of waveguide modes $t_{mn}(\nu)$ by means of a two-fold sine transformation:
\begin{align}
\begin{split}
    t_{mn} = \sum_{y_1,y_2} &t'(y_2,y_1) \sqrt{k_n(\nu) k_m(\nu)}\,\text{sin}\left(\frac{m\pi}{W}y_2\right) \text{sin}\left(\frac{n\pi}{W}y_1\right) .
\end{split}
\end{align}
\noindent Here, the transverse mode number $n$ is given by $k_n = (2\pi / c_0) \sqrt{\nu^2 - (n \nu_c)^2}$, the cut-off frequency is $\nu_c = c_0 / 2h$, $c_0$ is the speed of light and $h = 8$~mm is the height of the waveguide.\\\indent
The elements of $t^0(\nu)$ for an empty waveguide are presented in Fig.~\ref{fig:TM} at $\nu = 7$~GHz and $\nu = 11.2$~GHz. In both cases, the transmission matrix is diagonal as expected. This confirms that the coupling between our antennas is small and barely impacts our results. A strong coupling may indeed result in off-diagonal terms with high amplitude. The transmission of the last mode is seen to be slightly smaller than the transmission of the first modes. This results from losses within the waveguide as the last mode is associated with a larger time delay.
\begin{figure}[t!]
    \centering
    \includegraphics[width=\columnwidth]{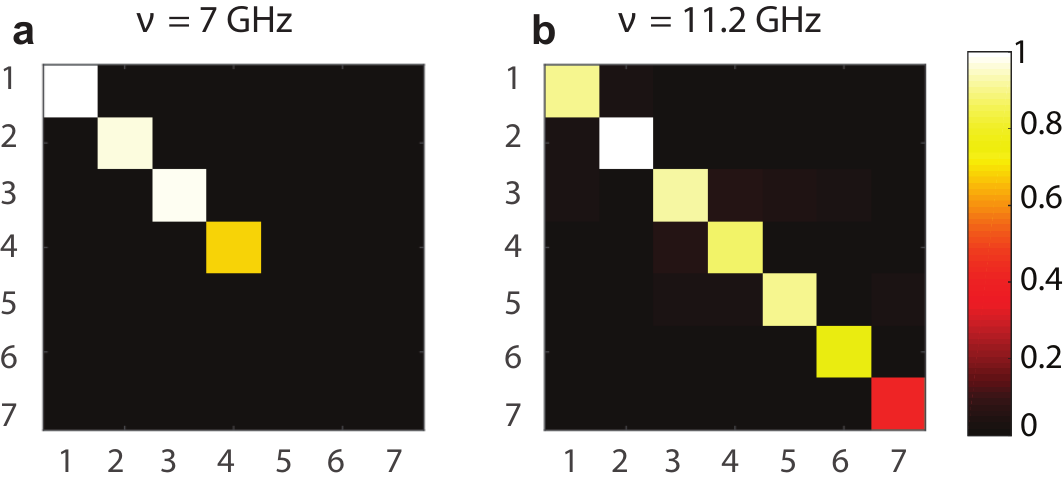}
    \caption{\new{Experimental intensity of the elements of the transmission $|t^0_{mn}|^2$ in the basis of waveguide modes for an empty waveguide at 7 and 11.2 GHz. At these frequencies, the waveguide supports $N=4$ and $N=7$ modes, respectively.}}
    \label{fig:TM}
\end{figure}
Because our antennas are weakly coupled to the waveguide, the transmission coefficients found from the two-fold sine transformation are not flux-normalized. As shown in the Methods section of the main text, we therefore normalize the transmission for each waveguide mode $n$, $T_n(\nu) = \Sigma_{m} |t_{mn}(\nu)|^2$, by its value for an empty waveguide, $T^0_n(\nu)$:
\begin{equation}
    {T}(\nu) = \frac{1}{N} \sum_{n=1}^N \frac{T_n(\nu)}{T_n^0(\nu)} .
\end{equation}
\noindent This is equivalent to normalizing the transmission matrix such as:
\begin{equation}
    {\Tilde{t}_{nm}}(\nu) = \frac{{{t}_{nm}}(\nu)}{\sqrt{T^0_n(\nu)}},
\end{equation}
and calculating the average transmission of each mode with $T(\nu) = [\Sigma_{n,m=1}^N|{\Tilde{t}_{nm}}(\nu)|^2]/N$.
\begin{figure}[t!]
    \centering
    \includegraphics[width=\columnwidth]{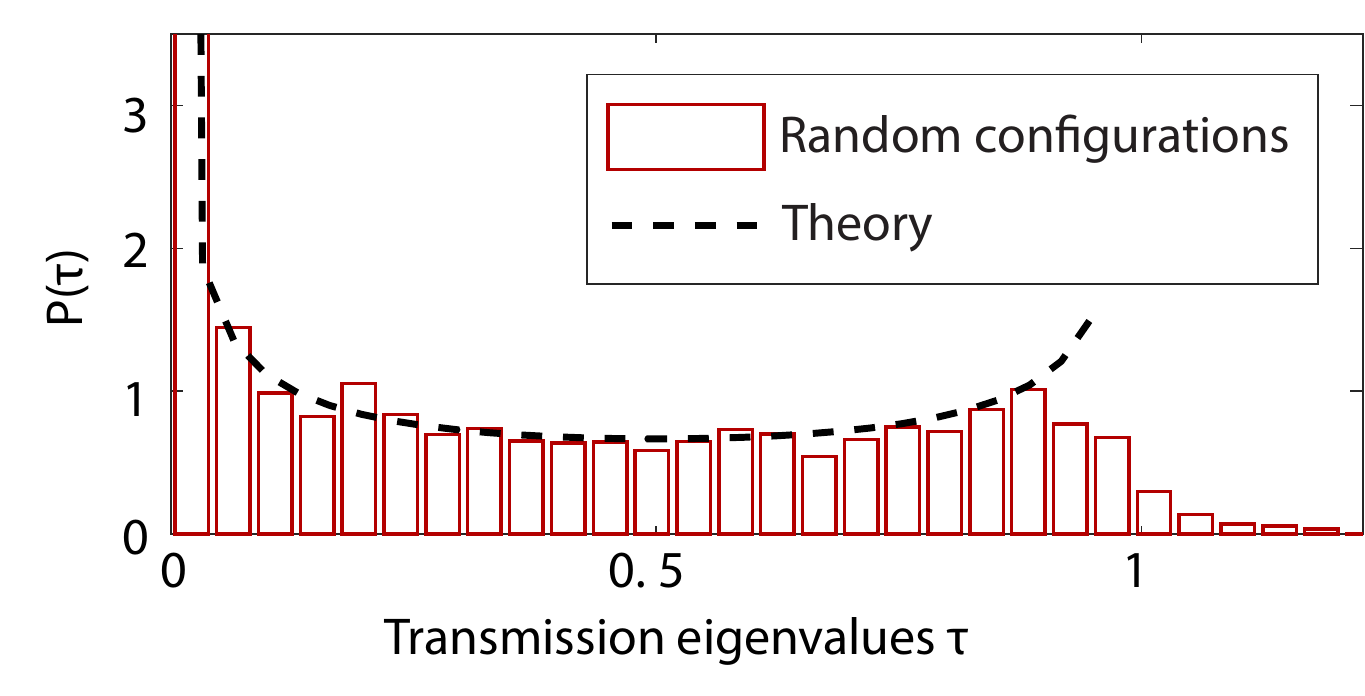}
    \caption{\new{Experimental transmission eigenvalue histogram for a waveguide supporting four modes compared to the bimodal law $P_0(\tau)$. The random disorders is composed of 6 aluminum cylinders and 34 Teflon cylinders.}}
    \label{fig:eigenvalues}
\end{figure}
To validate our experimental results, we first compute the distribution of transmission eigenvalues $\tau$ of ${\Tilde{t}}^\dagger (\nu){\Tilde{t}}(\nu)$ for waveguides with random disorder. Here the distribution is found from 10 random realizations with randomly located scatterers and an averaging over the frequency range [6.6-7.4]~GHz ($N=4$). The distribution is presented in Fig.~\ref{fig:eigenvalues}. As expected from diffusion theory, this distribution is bimodal with two peaks centered on closed channels with $\tau \sim 0$ and open channels with $\tau \sim 1$. The experimental result is in good agreement with the theoretical law:
\begin{equation}
    P_0(\tau) \propto \frac{1}{\tau \sqrt{1-\tau}}.
\end{equation}
However, in line with previous works on this subject \cite{gerardin_full_2014}, we observe the presence of transmission eigenvalues above unity. The bimodal distribution and more precisely the peak associated with open channels is indeed very sensitive to experimental noise. A nonunitarity of the scattering matrix due to experimental noise indeed leads to dramatic deviations from theory with transmission eigenvalues exhibiting coefficients larger than unity \cite{gerardin_full_2014}. These open channels virtually violate the energy conservation due to the noise level. As a result of the spreading of eigenvalues with large transmission, the amplitude of the corresponding peak also decreases. In our case, this noise level comes from the normalization of the elements of the transmission matrix using transmission through an empty sample. In particular, the last waveguide mode with a large angle between $\mathbf{k}$ and the longitudinal direction feature a large dwell time for the empty waveguide and is therefore very sensitive to global absorption. In the presence of weak disorder, the outgoing field is mixed in all modes leading to a possible smaller sensitivity to the absorption. As a result, the normalization can lead to transmission eigenvalues larger than unity.}

\begin{figure}[t!]
    \centering
    \includegraphics[width=\columnwidth]{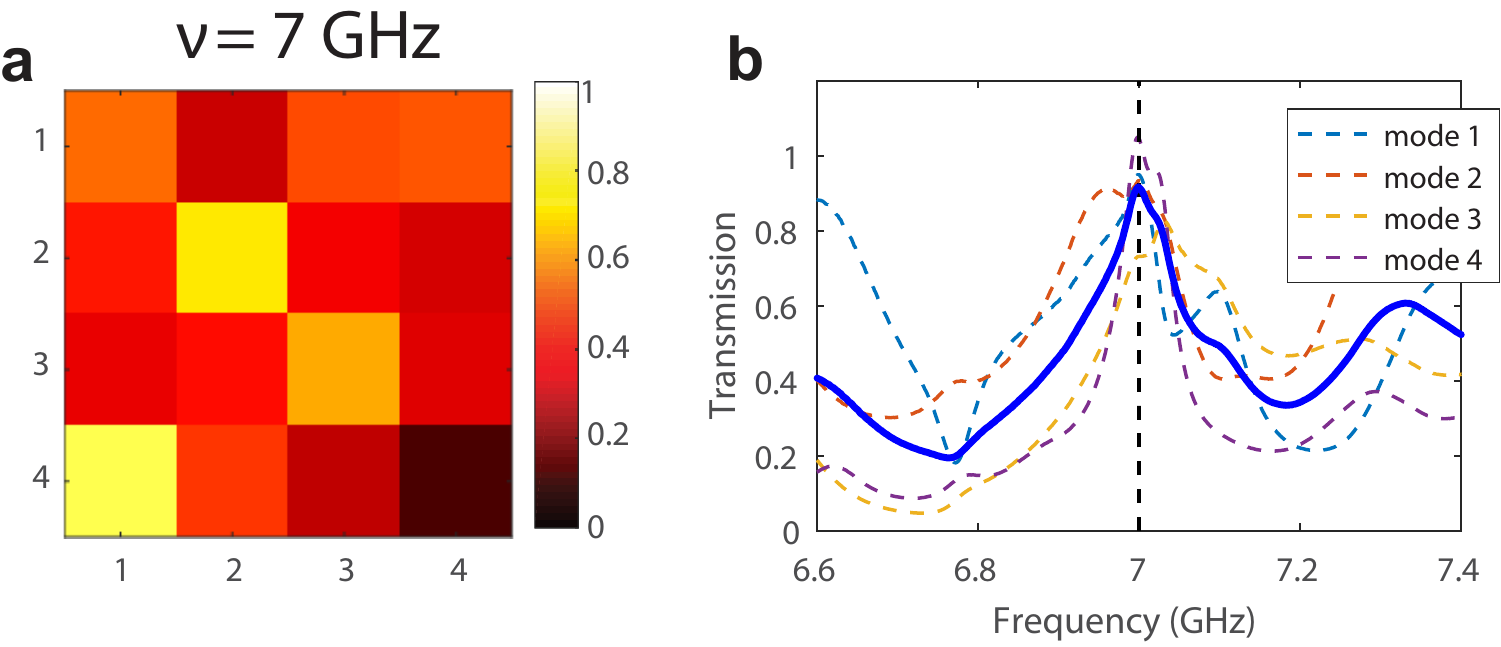}
    \caption{\new{\textbf{a}, Experimental intensity of the elements of the transmission $|t_{mn}|^2$ in the basis of waveguide modes at 7 GHz for a sample of complete transmission with 52 scatterers. \textbf{b}, Spectrum of the transmission of each mode through the waveguide (dotted line). The average transmission for the four modes is represented with the blue line.}}
    \label{fig:TM_completeT}
\end{figure}

\new{Experimentally, we implement anti-reflection structures by projecting on the waveguide an image of the scatterer positions found numerically using a video projector. The image is calibrated to minimize positioning errors. The cylinders are then placed manually. Small inaccuracies may result from this procedure but the overall agreement between numerical and experimental results is excellent as seen in Fig. 3 of the main text.}

\new{To further confirm our normalization procedure, we show in Fig.~\ref{fig:TM_completeT} the transmission associated to each incoming mode of the waveguide for a sample with complete transmission. The configuration corresponds to Fig.~3g of the main text. The transmission matrix $|{\Tilde{t}_{ij}}|^2$ is seen to be random in Fig.~\ref{fig:TM_completeT}a as a result of strong mode mixing. Nevertheless, each waveguide mode provides almost perfect transmission at $\nu_0 = 7$~GHz (see Fig.~\ref{fig:TM_completeT}b).}

\section{Complete transmission through a multichannel cavity}
\begin{figure}[t!]
    \centering
    \includegraphics[width=\columnwidth]{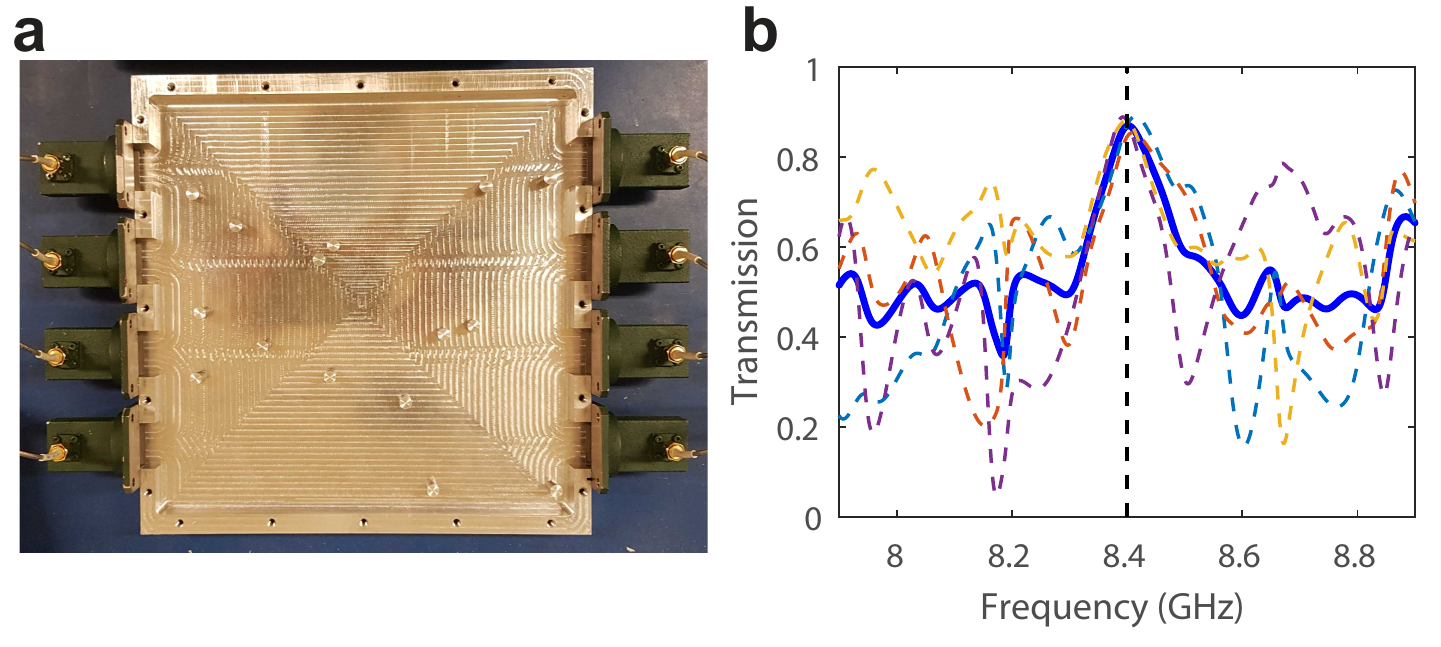}
    \caption{\new{\textbf{a}, Photography of the cavity. The top plate has been removed to see the interior of the cavity. Four transition-to-coax antennas are placed at the left and right side of the cavity. Measurement of the transmission matrix between these two arrays is carried out with a vector network analyzer. Fifteen metallic cylinders are placed at the positions determined numerically for perfect transmission. \textbf{b}, Total transmission $T_n(\nu) = \Sigma_m |t_{mn}(\nu)|^2$ for the four incoming channels (dashed lines) and its average $T(\nu) = (\Sigma_n T_n(\nu))/N $ over incoming channels (blue line). The placement of the cylinders correspond to positions optimised numerically for perfect transmission at $\nu_0 = 8.4$~GHz. Deviations from the maximal transmission value 1 is primarily due to absorption in the cavity.}}
    \label{fig:cavity}
\end{figure}

\new{To further illustrate the potential of our approach, we consider the case of a multichannel cavity. As shown in Fig.~\ref{fig:cavity}, the latter is a quasi-two dimensional square cavity of length and width $L=W=0.205$~m and height $h=0.010$~mm. A single vertically polarized mode can propagate within the cavity below $f = 14.7$~GHz. Two arrays of $N=4$ coax antennas are connected on the left and right interfaces. We carry out measurements of the $N\times N$ transmission matrix $t(\omega)$ between 7.8 and 9 GHz. Because coax-to-waveguides transitions are well-matched antennas between 7 and 12 GHz, the transmission coefficients are flux-normalized and no post-processing is needed. For an empty cavity, the average transmission $T(\omega)$ fluctuates within the selected frequency range between 0.37 and 0.83 numerically and between 0.3 and 0.7 experimentally as a consequence of absorption within the cavity.}

\new{We then gradually optimise the positions of 15 metallic cylinders of radius $r=3$~mm in numerical simulations to reach complete transmission at $\nu_0 = 8.4$~GHz. Because this configuration does not enable to write the complete scattering matrix as a composite expression of the scattering matrix of the empty cavity, the cost function $f=\mathrm{Tr} t^\dagger t /N$ is directly estimated in terms of the transmission matrix of the full system. The maximum transmission at the end of the numerical optimisation reaches 0.998. We then implement experimentally the numerical solution. The transmission spectrum nicely reproduces the numerical result with a maximal transmission of 0.9 at $\nu_0$. We observe an overall very good agreement between numerical simulations and experimental results even though maximal transmission is slightly reduced by the inevitable presence of absorption.}

\section{Experimental inaccuracies}

\begin{figure}[t!]
    \centering
    \includegraphics[width=\columnwidth]{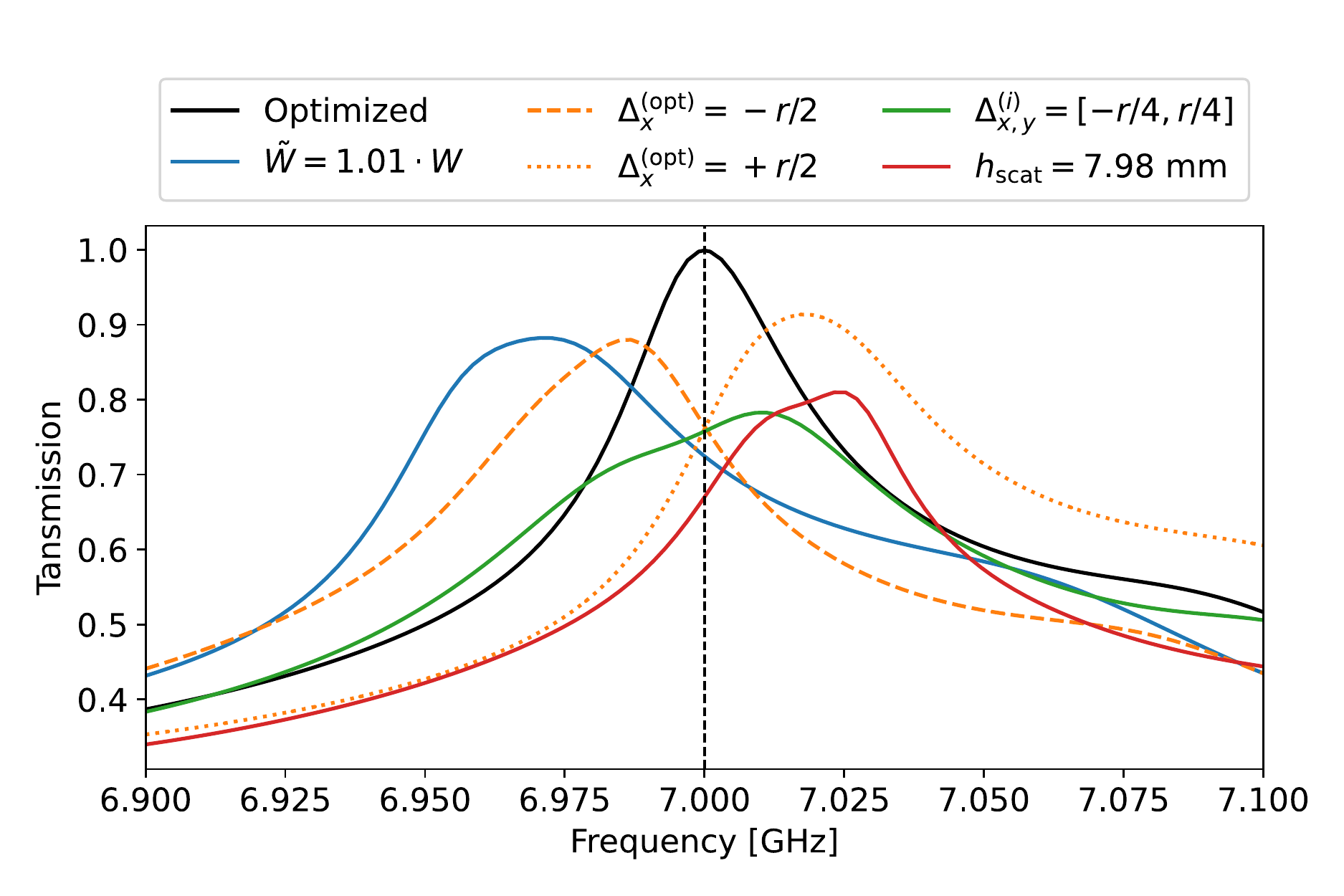}
    \caption{Transmission spectrum of the optimised sample with 52 scatterers (black solid line), where the vertical dashed line marks the frequency at which the optimisation has been performed. Increasing the waveguide width by 1\% of the initial width $W$ (blue solid line) causes the peak to shift to lower frequencies, where a global shift $\Delta_x^\mathrm{(opt)}$ in the negative/positive longitudinal direction of half a scatterer radius $r$ of only the optimised scatterers results in a shift to lower/higher frequencies (orange dashed/dotted line). In both cases, the peaks are also lowered due to the deviation from the optimised configuration. Applying small random displacements $\Delta_{x,y}^\mathrm{(i)}$ in the range $[-r/4,r/4]$ in $x$- and $y$-direction to every single scatterer also result in a reduction of the peak height and a shift (green solid line). Performing full vectorial 3D simulations, we also find that using cylindrical scatterers with a height $h_\mathrm{scat} = 7.98 \ \mathrm{mm}$ smaller than the waveguide height $h = 8  \ \mathrm{mm}$ also lowers and shifts the peak to higher frequencies (red solid line).}
    \label{fig:figa5}
\end{figure}

To understand the small shifts of the transmission peaks found in the experimental measurements (see Fig.~3f of the main text for complete transmission and Fig.~3e,f,h for maximal reflection) with respect to the ones found via the numerical optimisations (see Fig.~3a,b of the main text), we perform numerical simulations under more realistic conditions including absorption or possible perturbations of the numerically optimised configurations in the perfect waveguide featuring full transmission. 

As in every experimental setup, uniform absorption is typically present and affects the transmission spectrum. Thus we add a uniform imaginary part of $n_I = 3 \times 10^{-4}$ to the refractive index distribution of the waveguide containing the optimised configuration and as shown in Fig.~\ref{fig:figa4} this only lowers the transmission peak, but doesn’t shift it noticeably. 

We first consider uncertainties in the experimental placement of the scatterers, such as when the experimental scatterer positions are slightly different from the ones obtained from the numerical optimisation. We investigate the effect of a small global shift of only the optimised part of the scattering configuration in the negative/positive longitudinal direction which causes a shift of the transmission peak to lower/higher frequencies, where the peak is typically also lowered (orange dashed/dotted line in Fig.~\ref{fig:figa5}). Since the uncertainties in the scatterer placement might not be global but rather random, we also investigated the effect of small random displacements of the optimised scatterer positions which causes peak shifts, lowerings and broadenings depending on the magnitude of the displacements (green solid line in Fig.~\ref{fig:figa5}). 

Because of possible fabrication uncertainties and the skin effect in the metallic waveguide  walls, we also consider slightly different waveguide dimensions. Specifically, we study the effect of a slightly wider waveguide (with the scatterers kept transversally in the middle of the waveguide), which lowers the transmission peak and shifts it to lower frequencies (blue solid line in Fig.~\ref{fig:figa5}). 

Moreover, in the experiment not all scatterers may reach the waveguide’s top plate perfectly. The resulting gap can then cause the waves to scatterer off the top edge of the cylindrical scatterers exciting evanescent modes, which might change the transmission spectrum due to coupling to the surrounding scatterers. To examine this effect, we perform 3D simulations in which we solve the vectorial Helmholtz equation $\nabla \times \nabla \times \mathbf{E}(\mathbf{r}) - n^2(\mathbf{r}) k_0^2 \mathbf{E}(\mathbf{r}) = 0$. We find that a small gap (especially for  metallic scatterers) results in lowered transmission peak which is shifted to higher frequencies (red solid line in Fig.~\ref{fig:figa5}).

In the experiment, we most likely observe a combination of all these effects. Additionally, the experiment can suffer from spurious reflections at the non-perfect absorbers at the waveguide ends as well as from scattering off the antennas used to inject and measure the waves, which can further effect the transmission spectrum.

\bibliographystyle{apsrev4-1}
\providecommand{\noopsort}[1]{}\providecommand{\singleletter}[1]{#1}%

\end{document}